\documentclass[sn-mathphys-ay,pdflatex]{sn-jnl}% Math and Physical Sciences Author Year Reference Style

\usepackage{graphicx}%
\usepackage{multirow}%
\usepackage{amsmath,amssymb,amsfonts}%
\usepackage{amsthm}%
\usepackage{mathrsfs}%
\usepackage[title]{appendix}%
\usepackage{xcolor}%
\usepackage{textcomp}%
\usepackage{manyfoot}%
\usepackage{booktabs}%
\usepackage{algorithm}%
\usepackage{algorithmicx}%
\usepackage{algpseudocode}%
\usepackage{listings}%
\usepackage{soul}%
\usepackage{textalpha}
\usepackage{gensymb}
\newcommand{\sarc}{$^{\prime\prime}\!\!$}

\graphicspath{{./figures/}}
\begin{document}

\title[Sub-arcsec imaging with the ILT]{A decade of sub-arcsecond imaging with the International LOFAR Telescope}

\author*[1,2]{\fnm{Leah K.} \sur{Morabito}}\email{leah.k.morabito@durham.ac.uk} 

\author[3]{\fnm{Neal} \sur{Jackson}}\email{neal.jackson@manchester.ac.uk}

\author[4]{\fnm{Jurjen} \sur{de Jong}}\email{jurjendejong@strw.leidenuniv.nl}

\author[1]{\fnm{Emmy} \sur{Escott}}\email{emily.l.escott@durham.ac.uk}

\author[4,5]{\fnm{Christian} \sur{Groeneveld}}\email{groeneveld@strw.leidenuniv.nl}

\author[6,7]{\fnm{Vijay} \sur{Mahatma}}\email{vm462@cam.ac.uk}

\author[1,4]{\fnm{James} \sur{Petley}}\email{petley@strw.leidenuniv.nl}

\author[1]{\fnm{Frits} \sur{Sweijen}}\email{frits.sweijen@durham.ac.uk}

\author[1,2]{\fnm{Roland} \sur{Timmerman}}\email{roland.timmerman@durham.ac.uk}

\author[4]{\fnm{Reinout J.} \sur{van Weeren}}\email{rvweeren@strw.leidenuniv.nl}


\affil[1]{Centre for Extragalactic Astronomy, \orgdiv{Department of Physics}, \orgname{Durham University}, \orgaddress{\street{South Road}, \city{Durham}, \postcode{DH1 3LE}, \country{United Kingdom}}}

\affil[2]{Institute for Computational Cosmology, \orgdiv{Department of Physics}, \orgname{Durham University}, \orgaddress{\street{South Road}, \city{Durham}, \postcode{DH1 3LE}, \country{United Kingdom}}}

\affil[3]{\orgname{University of Manchester}, \orgdiv{Jodrell Bank Centre for Astrophysics, Department of Physics and Astronomy}, \orgaddress{\street{Oxford Rd}, \city{Manchester} \postcode{M13 9PL}, \country{United Kingdom}}}

\affil[4]{\orgdiv{Leiden Observatory}, \orgname{Leiden University}, \orgaddress{\street{PO Box 9513}, \postcode{2300 RA} \city{Leiden}, \country{The Netherlands}}}

\affil[5]{\orgname{INAF - Instituto di Radioastronomia}, \orgaddress{\street{via P. Gobetti 101}, \postcode{40129} \city{Bologna}, \country{Italy}}}

\affil[6]{\orgdiv{Cavendish Laboratory- Astrophysics Group}, \orgname{University of Cambridge}, \orgaddress{\street{19 JJ Thomson Avenue}, \city{Cambridge}, \postcode{CB3 0HE}, \country{United Kingdom}}}

\affil[7]{\orgdiv{Kavli Institute for Cosmology}, \orgname{University of Cambridge}, \orgaddress{\street{Madingley Road}, \city{Cambridge}, \postcode{CB3 0HA}, \country{United Kingdom}}}


\abstract{The International LOFAR Telescope (ILT) is a pan-European radio interferometer with baselines up to 2,000 km. This provides sub-arcsecond resolution at frequencies of $<$200 MHz. Since starting science operations in 2012, the ILT has carried out observations for the state-of-the-art LOFAR Two-metre Sky Survey, which has 6 arcsec resolution at 144 MHz. Wide-area surveys at low frequencies, while scientifically productive, have to compromise on resolution. Sub-arcsecond imaging with the ILT has become more accessible over the last decade, thanks to efforts to build a publicly available pipeline using LOFAR-specific tools, which has resulted in a dramatic increase in the number of publications. The ILT's combination of resolution, field of view, and low observing frequency make it a unique instrument for a wide range of scientific applications, and it will remain unparalleled even in the era of the Square Kilometre Array Observatory. Here we provide an overview of the technical considerations and calibration methods sub-arcsecond imaging with the ILT. This is followed by a review of the unique capabilities unlocked by sub-arcsecond imaging with the ILT, using examples from the literature for demonstration. Finally we describe ongoing work including: surveying large areas of the sky at high resolution, going deeper in fields with excellent ancillary information, producing images of polarisation, and extending to lower frequencies ($<$100 MHz). }


\maketitle

\section{Introduction}
\label{sec:intro}

Radio surveys of extragalactic sources have led to transformational science in the field of active galactic nuclei (AGN) and galaxy evolution, starting with the discovery of quasars in the 1960s \citep[][]{schmidt_3c_1963,oke_absolute_1963}. Surveys in particular enable statistical studies to investigate trends in populations, and if the studies are large enough, it is possible to control for factors like stellar mass and redshift. Surveys like the NRAO VLA Sky Survey \citep[NVSS;][]{condon_nrao_1998} and the Faint Images of the Radio Sky at Twenty-cm (FIRST) survey \citep{becker_first_1995} have been used widely by the astronomical community over the past several decades, resulting collectively in $\sim$7,000 papers on the topic of galaxies and/or galaxy evolution. In the past decade, attention has turned to deeper surveys like the LOw Frequency ARray \citep[LOFAR;][]{van_haarlem_lofar:_2013} Two-metre Sky Survey \citep[LoTSS;][]{shimwell_lofar_2019,shimwell_lofar_2022} which is now the largest ever radio survey (by source count and optical identifications) at 4.4 million sources from less than a third of the final survey area (the Northern sky). With over 1,200 citations in just over five years, LoTSS is one of the most productive radio surveys of all time. 

Large-area radio surveys compromise on sensitivity and/or resolution to achieve higher survey efficiency. Radio observing time is a finite resource, and survey efficiency increases with reducing the required observation time per square degree. Generally this means compromising on sensitivity to cover larger areas, and/or using arrays and configurations with wider fields of view, which generally come with lower resolution. This means we often have to rely on information from other wavebands to help characterise the radio emission from a source. The classic way to do this is using radio loudness, $R$ \citep[canonically defined as the ratio between 5$\,$GHz and $B$-band (optical) luminosities;][]{kellermann_vla_1989}. Radio loud sources, historically defined as $R\gtrsim 10$ at 5$\,$GHz, have been well-studied and we know that their radio emission is dominated by powerful relativistic jets which can extend up to Mpc in scale \citep[e.g.,][]{oei_black_2024,dabhade_giant_2020,willis_3c236_1974}. The distribution of $R$ from surveys like FIRST and LoTSS make it clear that there is also a population with low $R$, with the addition of a very long `tail' of sources with high $R$ \citep[see, e.g.,][]{white_signals_2007,gurkan_lotsshetdex_2019}. LoTSS is about an order of magnitude deeper than FIRST\footnote{for a source with a canonical spectral index $\alpha=-0.7$, where $S\propto\nu^{\alpha}$} which has opened up studies into the radio-quiet AGN population, where the source of radio emission is still unknown \citep[see][and references therein]{panessa_origin_2019}. Most of these radio-quiet sources are unresolved in wide-area surveys, often limiting the observable information to only total flux density. With sufficient frequency coverage (e.g., cross-matching FIRST and LoTSS) we can extend this to radio spectral index, but the varying sensitivity of different surveys makes it difficult to build complete samples to understand the intrinsic population of radio sources. Resolution is the final key to this puzzle, to add information on the morphology of radio sources.

\begin{figure}
    \centering
    \includegraphics[width=\linewidth,clip,trim=10cm 4cm 8cm 6cm]{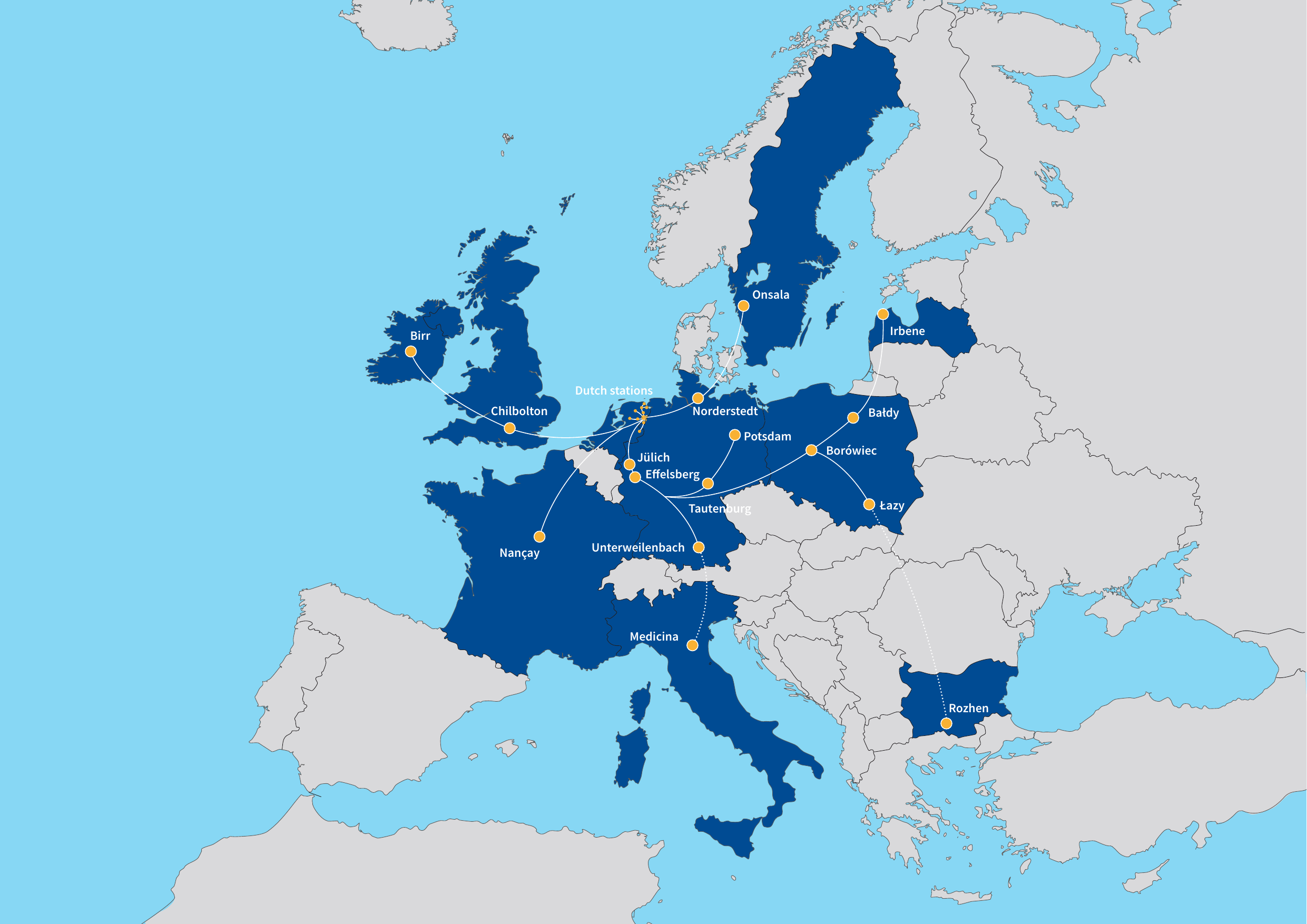}
    \caption{The locations of current and planned ILT stations. There are currently 52 stations, with two stations which will join the array in the near future, one each in Italy (Medicina) and Bulgaria (Rozhen). Image credit: ASTRON, Netherlands Institute for Radio Astronomy.}
    \label{fig:lofar}
\end{figure}

Sub-arcsecond resolution is possible with the International LOFAR Telescope (ILT), which currently spans 8 European countries with baselines up to $\sim$2,000 km, see Fig.~\ref{fig:lofar}. This provides a resolution of 0.3 arcsec at 144 MHz, compared to the 6 arcsec resolution of LoTSS using only the Dutch stations. Early studies with the full ILT utilised custom-built data processing strategies, relying on traditional very long baseline interferometry (VLBI) techniques like fringe-fitting \citep{varenius_subarcsecond_2015,varenius_subarcsecond_2016,morabito_lofar_2016,ramirez-olivencia_sub-arcsecond_2018,harris_lofar_2019}, yielding spectacular results on nearby star-forming galaxies and more distant AGN. These early studies were crucial for understanding the calibration challenges of sub-arcsecond imaging with the ILT. 

Pulling together the expert community knowledge, \cite{morabito_sub-arcsecond_2022} defined the standard calibration strategy, using LOFAR-specific tools to make the data processing easier. This coincided with a public release of the pipeline and the release of the completed Long Baseline Calibrator Survey \citep[LBCS;][]{jackson_sub-arcsecond_2022}. A coordinated effort using this pipeline resulted in a special issue of Astronomy \& Astrophysics, which more than doubled the number of publications using high-resolution ILT imaging. These studies again focused on individual objects, looking at star formation and AGN in nearby galaxies, and spatially-resolved AGN emission in distant galaxies. The community has continued to build on and further develop the pipeline, including the first full field-of-view image at sub-arcsecond resolution with the ILT \citep{sweijen_deep_2022}. This covers an area of 6.6 degree$^2$ thanks to the ILT's large field of view. \cite{de_jong_into_2024} has followed this up with imaging at 0.3, 0.6, and 1.2 arcsec imaging of the ELAIS-N1 field \citep[in addition to the already published 6 arcsec resolution images,][]{sabater_lofar_2021}. This represents the current state-of-the-art high-resolution imaging with the ILT. 

In this review, we outline the methods in Section~\ref{sec:methods}, with a focus on the technical considerations not covered in other publications. This is followed by an overview of the advantages of sub-arcsecond imaging with the ILT in Section~\ref{sec:unique}, using demonstrations from the literature to illustrate the unique capabilities. We finish with an overview of ongoing work and future plans in Section~\ref{sec:ongoing}, and conclude in Section~\ref{sec:conclusions}. 

\section{Technical considerations and methods}
\label{sec:methods}
The calibration strategies for high-resolution imaging with the ILT are described in other publications, which are cited below when relevant. Here we discuss several technical considerations which, although important, are not be covered in detail elsewhere. Following this we summarise the overall calibration and imaging approach. 

\subsection{Technical considerations}
\label{subsec:tech}
Calibration and imaging with the full ILT brings a few caveats along. While some of these also impact other VLBI arrays, there are some unique challenges for the ILT which we outline here. 

\begin{figure}
    \centering
    \includegraphics[width=0.5\linewidth]{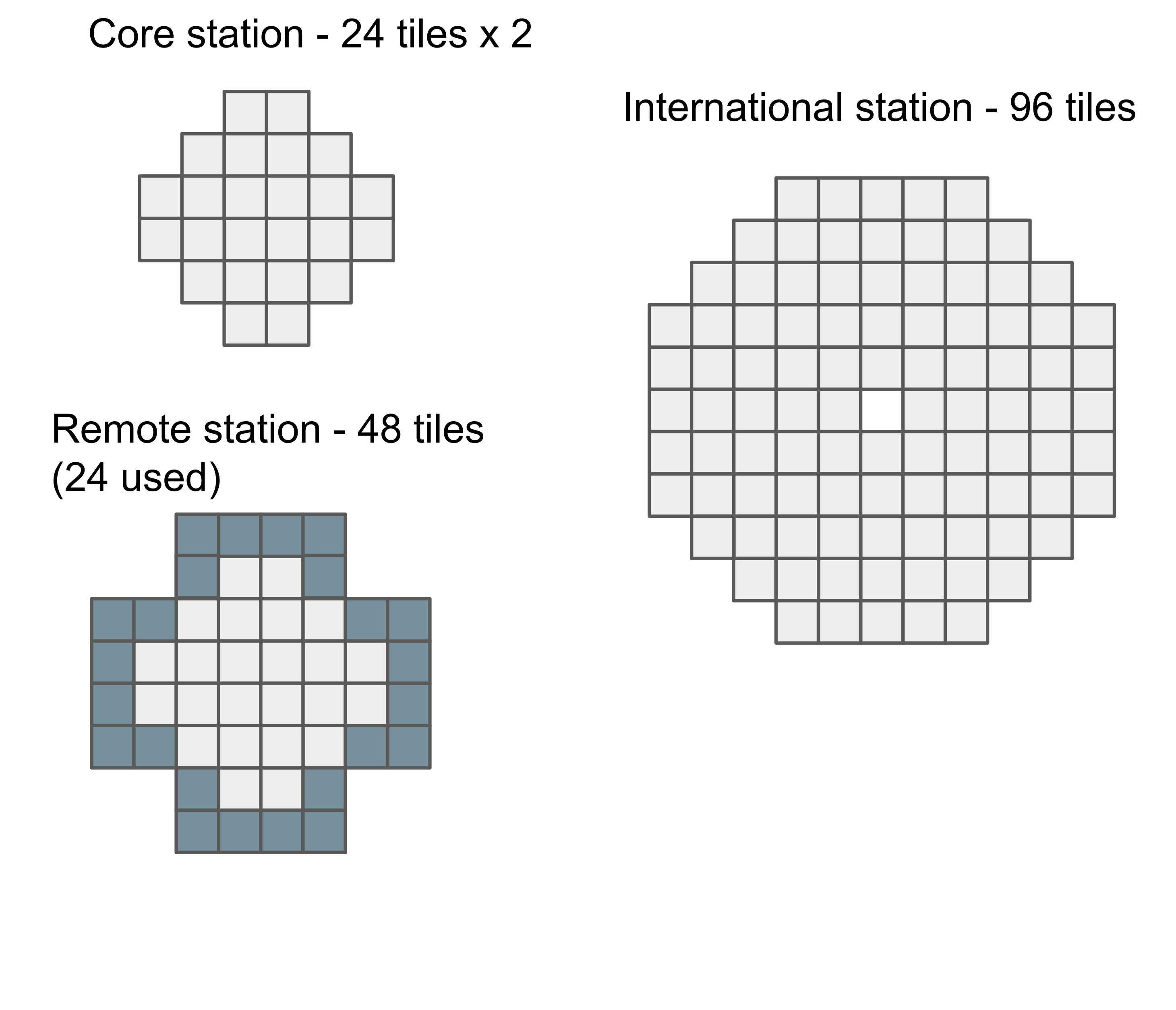}
    \includegraphics[width=0.48\linewidth]{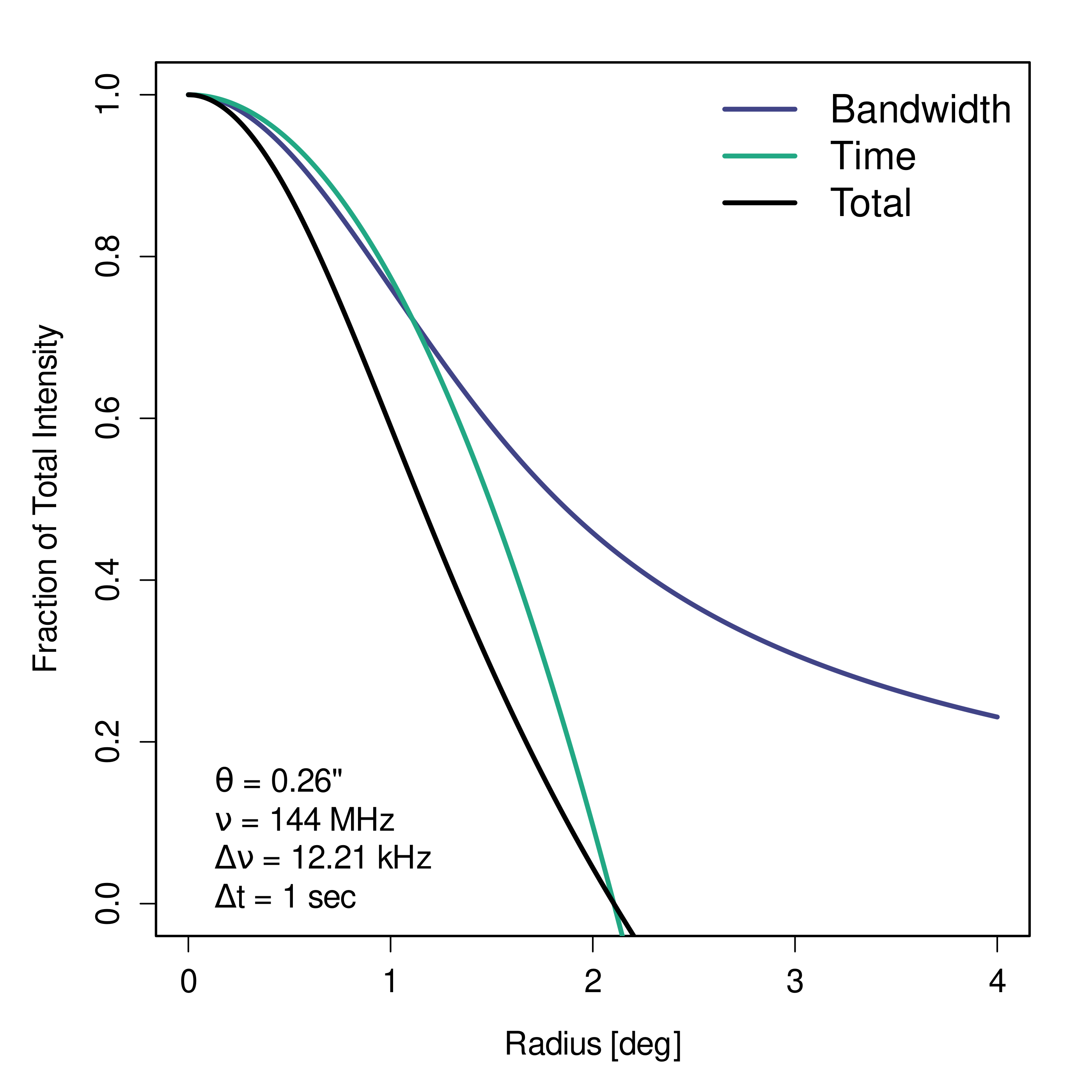}
    \caption{\textit{Left:} The tile layouts for the High Band Antenna (HBA) stations, showing their relative physical size. Although the Remote stations have 48 tiles, standard operations use only the inner 24 tiles to match the Core stations. The international stations by default use all 96 tiles. \textit{Right:} The theoretical intensity losses from smearing from frequency and time averaging, and their combination, for the longest baseline of 1980 km (Ireland to Poland). The curves shown are for the standard time and frequency averaging in LOFAR Two-Metre Sky Survey (LoTSS) datasets.}
    \label{fig:smearing}
\end{figure}

\paragraph{Mis-matched fields of view} The Dutch array uses a fairly homogenous configuration for a standard LoTSS observation, where the $48$-tile remote stations only use the inner $24$ tiles such that they are matched with the $24$-tile core stations. This yields an effective station diameter of approximately $30.75\mathrm{m}$, for all the Dutch stations, in a typical LoTSS observation. The international stations, however, consist of $96$ tiles and are used in full as station sensitivity becomes increasingly important on longer baselines.  With a diameter of $56.5\ \mathrm{m}$, they are almost twice the size of the Dutch stations\footnote{\url{https://science.astron.nl/telescopes/lofar/lofar-system-overview/observing-modes/lofar-imaging-capabilities-and-sensitivity/}}, shrinking their field of view by an equal amount. The relative sizes of the high band antenna (HBA) tiles are shown in the left panel of Figure~\ref{fig:smearing}. At the typical central frequency of the HBA of $144\ \mathrm{MHz}$, the full-width half-maximum (FWHM) of the primary beam is $3.9^\circ$ for the Dutch stations, but only $2.1^\circ$ for the international stations. In order to address this, care must be taken to remove sources outside the international station FWHM from the data using a model from, for example, LoTSS, before processing the data for full field-of-view (FoV) imaging.

\paragraph{Time and bandwidth smearing}
Time and bandwidth smearing manifest in radio images due to finite temporal and spectral sampling of the visibilities. The strength of both of these effects increases with distance from the phase centre, limiting the usable field of view. Time smearing causes an azimuthal blurring around the phase centre. Bandwidth smearing is a form of chromatic aberration that produces radial blurring pointing towards the phase centre. Visually these aberrations only show up in extreme cases. In general, they manifest more subtly as a reduction in the observed peak intensity of a source and a size that is larger than one would expect from the point spread function. For ILT observations, data are stored at a time resolution of $1\ \mathrm{s}$ and a frequency resolution of $12.207\ \mathrm{kHz}$. 
For interferometric observations this is the smallest time interval which can be recorded by the correlator. The frequency averaging was chosen so the smearing approximately matches the time smearing; this is a balance between not causing further smearing while reducing the data volume. We are careful to preserve this averaging as much as possible to mitigate smearing in our final images. For calibration of individual sources (i.e., in-field calibrators), the data are phase-shifted first and then can be safely averaged to larger time and frequency intervals, reducing the impact of smearing. 
The right panel of Figure~\ref{fig:smearing} shows the theoretical intensity losses due to these two effects for this level of data averaging at a frequency of 144 MHz for the ILT's longest baseline of $1980\ \mathrm{km}$, using the relevant equations from \cite{bridle_bandwidth_1999}. As can be seen, on the longest baseline(s) smearing has a significant impact.

\begin{figure}
    \centering
    \includegraphics[width=0.49\linewidth]{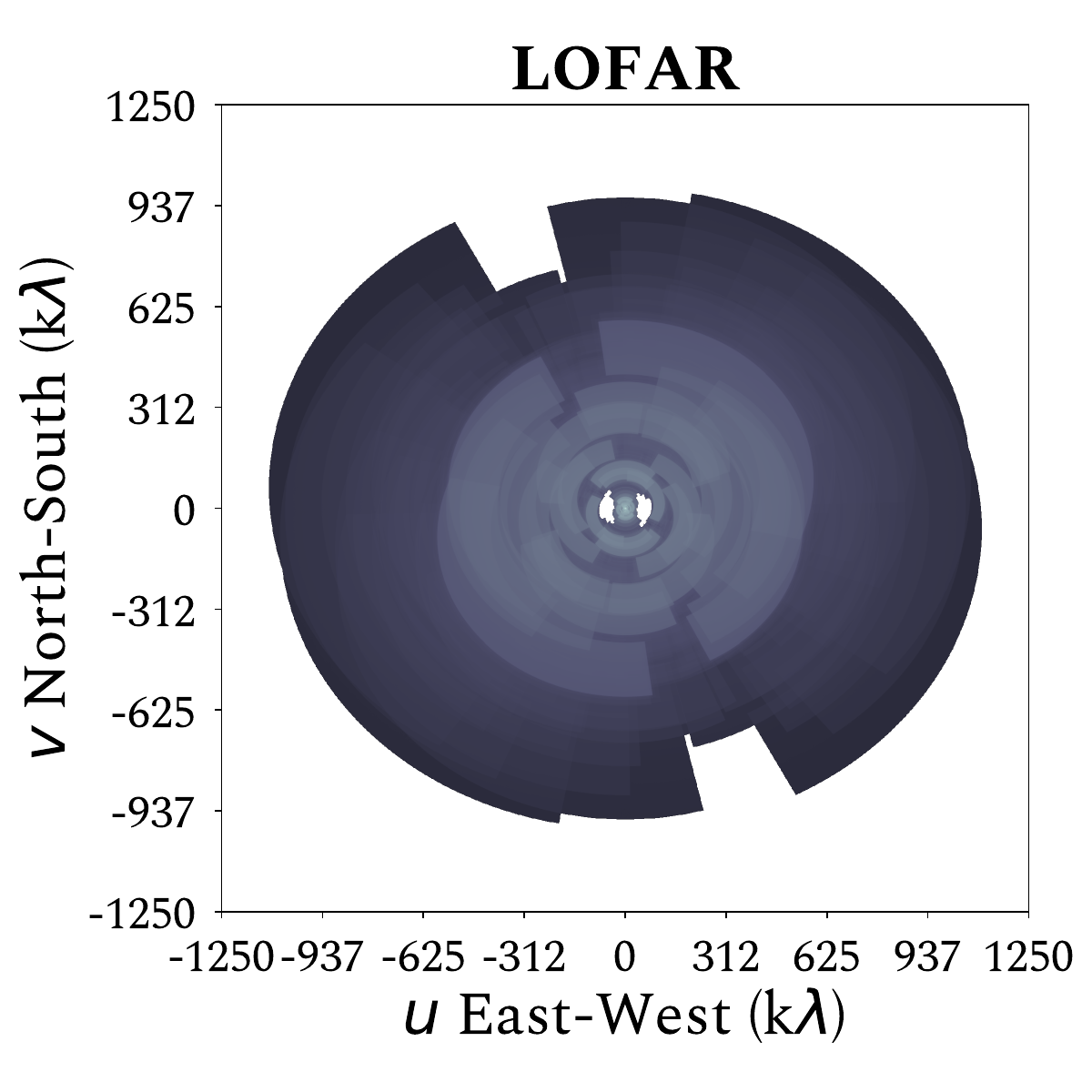}
    \includegraphics[width=0.49\linewidth]{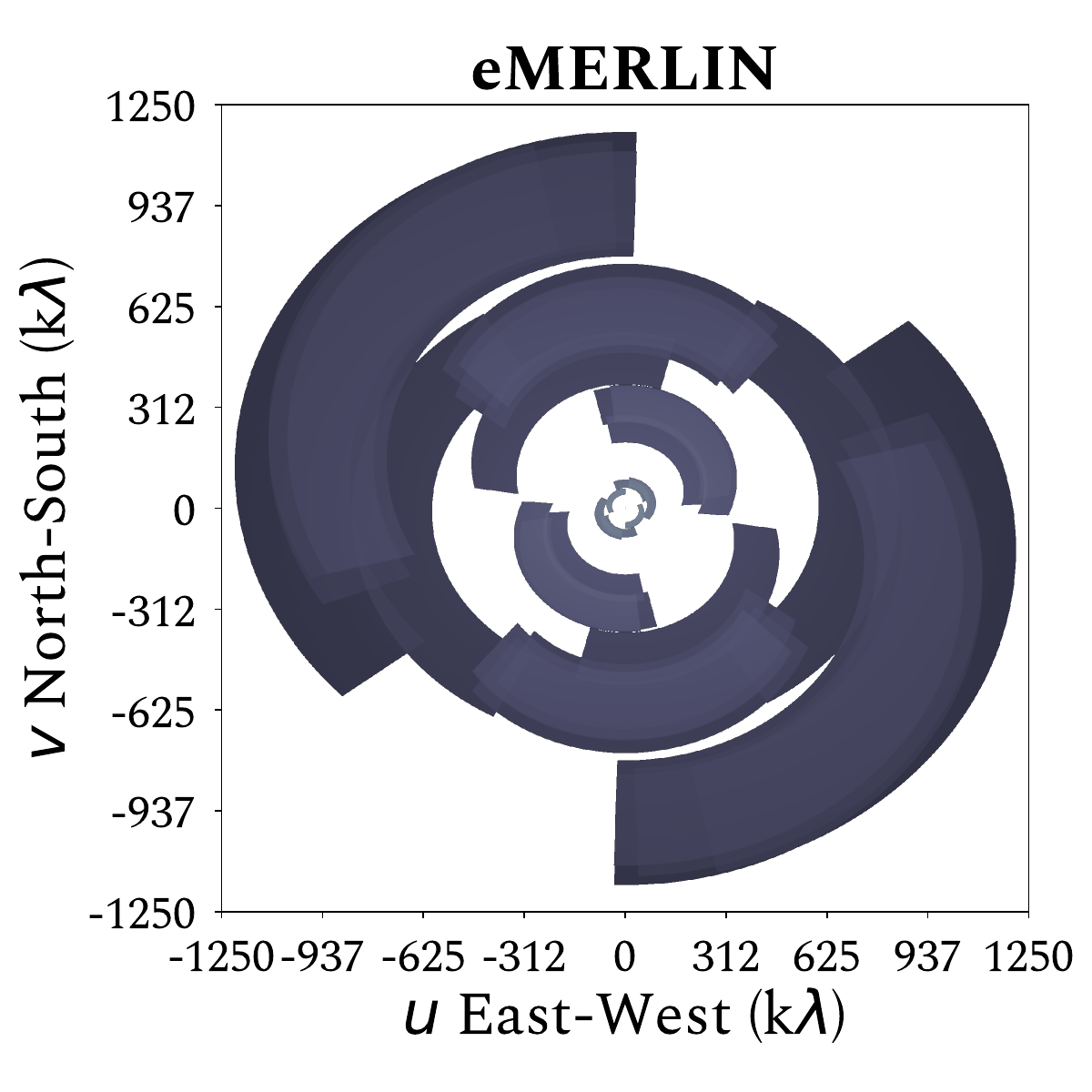}\\
    \includegraphics[width=0.49\linewidth]{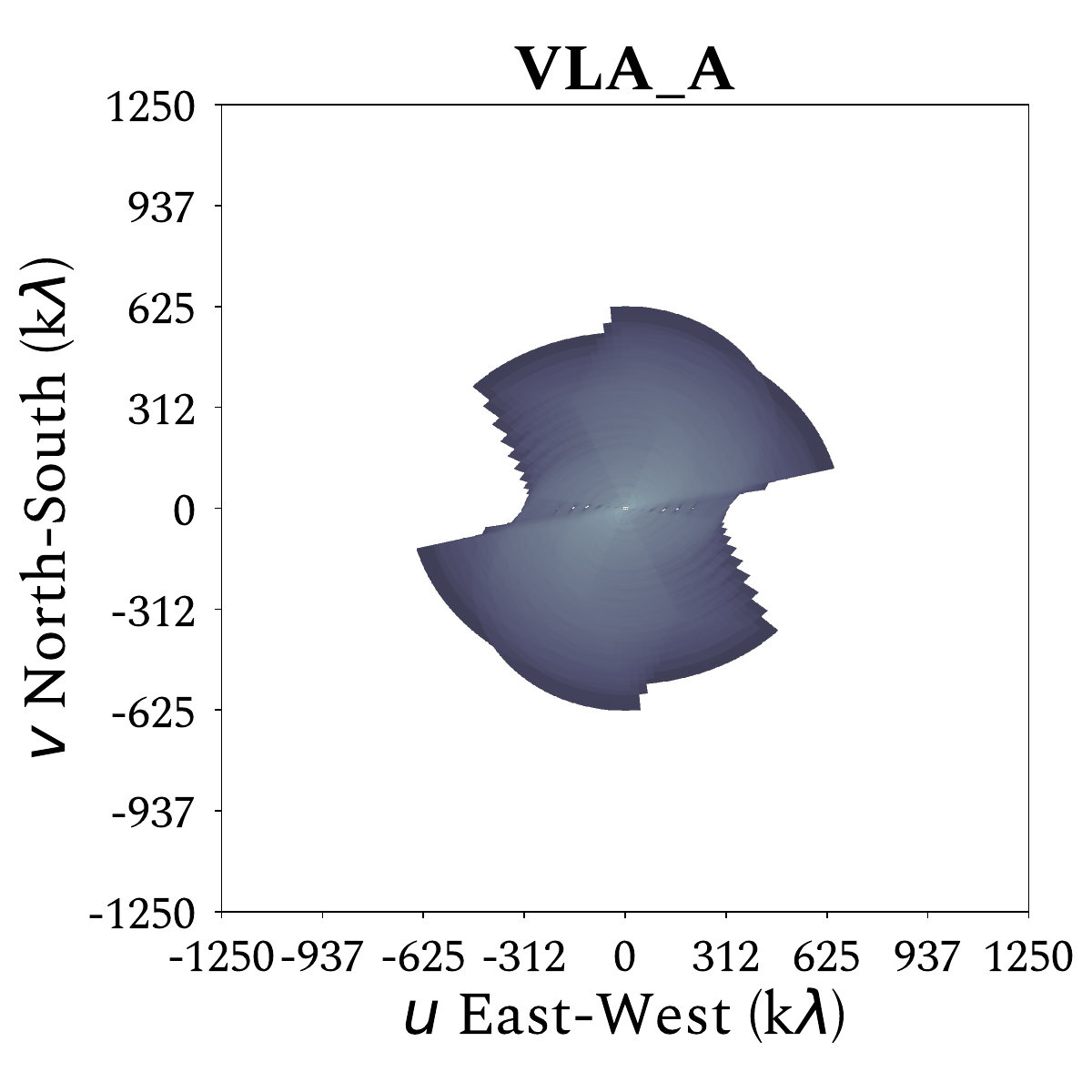}
    \caption{Comparison of the density in $uv$ coverage obtained from an $8$ hour observation with the International LOFAR Telescope ($120-168\ \mathrm{MHz}$, $\Delta\nu = 12.2\ \mathrm{kHz}$; \textit{left}), e-MERLIN at L-band ($1.2-1.7\ \mathrm{GHz}$, $\Delta\nu = 125\ \mathrm{kHz}$; \textit{middle}) and the Very Large Array (VLA) in A-configuration at C-band ($4.5-5.5\ \mathrm{GHz}$, $\Delta\nu = 2\ \mathrm{MHz}$; \textit{right}). All panels use an integration time of $1\ \mathrm{s}$, the same $uv$ cells and are normalised to unity with respect to each of their maximum densities. The observations were simulated for a declination of $\delta = 58\degree$ around transit, at an hour-angle range between $-4$ and $+4$ hours.}
    \label{fig:uvcov}
\end{figure}

\paragraph{$u$-$v$ coverage} One of the ILT's unique aspects is its dense coverage of the $u$-$v$ plane. This pushes it away from classical VLBI and more towards high-resolution interferometry. A typical eight-hour LoTSS observation fills almost the entire $u$-$v$ plane from a few $\lambda$ out to $\sim 1\,$M$\lambda$. Only a small gap on intermediate angular scales remains. Fig.~\ref{fig:uvcov} illustrates the $u$-$v$ coverage of a typical ILT observation over 8 hours and $48\,$MHz of bandwidth at HBA frequencies. It shows both the incredibly dense coverage, as well as how it provides an excellent match to other higher-frequency facilities such as the Multi-Element Radio Linked Interferometer Network (e-MERLIN) or the Very Large Array (VLA).

\begin{figure}
    \centering
    \includegraphics[width=0.8\linewidth]{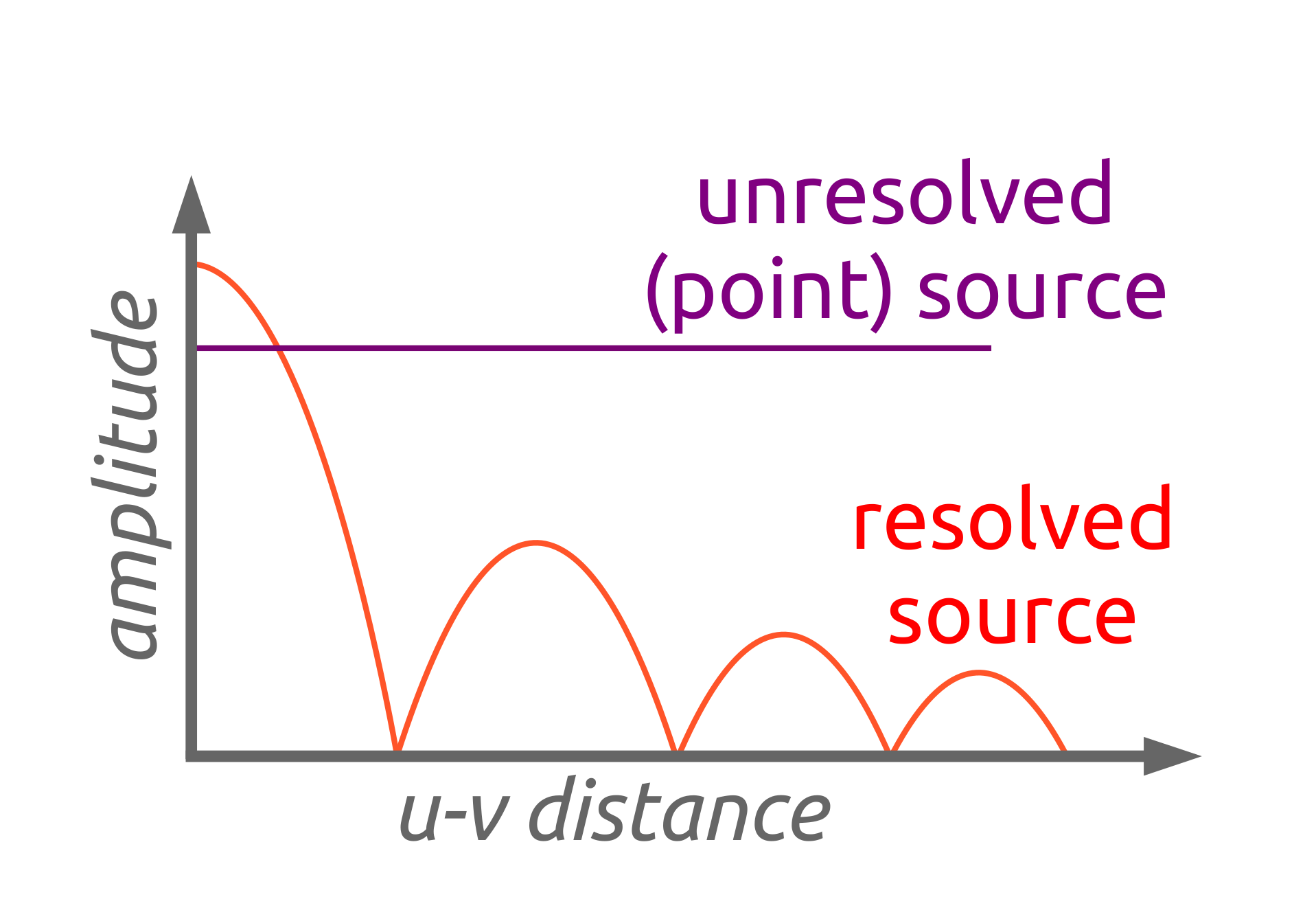}
    \caption{A cartoon illustration showing why high resolution reduces the signal to noise ratio on long baselines.  For a perfect point source, its amplitude remains constant at all baseline lengths (in units of $u$-$v$ distance), while the signal for a resolved source drops for longer baselines (larger $u$-$v$ distance).}
    \label{fig:ressnr}
\end{figure}

\paragraph{Lower signal to noise ratios for calibration}
The calibration of the ILT requires bright compact emission in a source. Compared to what the Dutch part of the array can achieve, the international baselines can improve upon the angular resolution of the observation twenty-fold. At these higher angular resolution, the majority of the observed emission tends to become resolved. This reduces the amount of signal available for the calibration of the international baselines, and thus reduces the overall signal to noise ratio available for calibration. Figure~\ref{fig:ressnr} provides an illustration. The LBCS found that, on average, the sky density of sources with enough signal to noise on which to calibrate is around 1 per square degree for the HBA. This is enough to be able to calibrate in any direction in the sky, as there will always be $\sim$5 good calibrator candidates in any field of view. 

\begin{figure}
    \centering
    \includegraphics[width=0.5\linewidth]{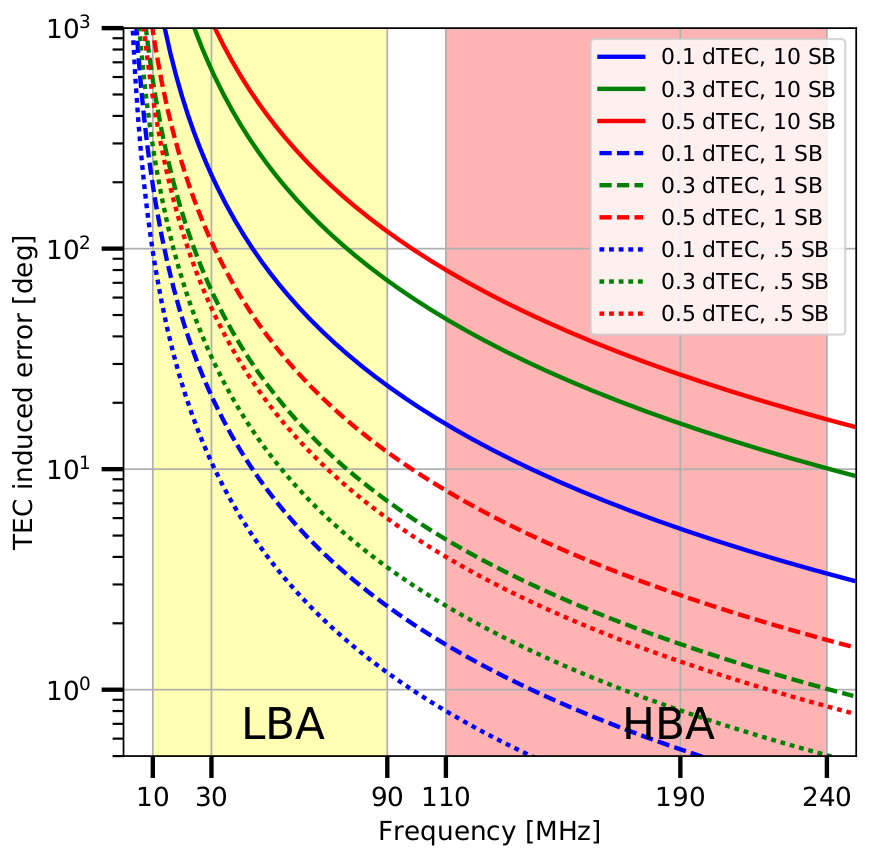}
    \caption{The phase error induced by differential Total Electron Content (dTEC), as a function of frequency. The errors increase drastically towards lower frequencies and for higher values of TEC, and when the data are averaged over larger numbers of subbands (SB; i.e. bandwidth). Figure credit: de Gasperin et al., A\&A, 615, A179, 2018, reproduced with permission \textcopyright\  ESO.}
    \label{fig:tec}
\end{figure}

\paragraph{The ionosphere}
The ionosphere introduces dispersive delays, which have a non-linear dependence on frequency. We characterise this by the Total Electron Content (TEC), where the phase $\phi$ is $\propto$ TEC$\,\nu^{-1}$. For the same TEC, the impact on phase therefore becomes larger at lower frequencies (smaller values of $\nu$). This is shown in Figure~\ref{fig:tec}, where the error induced in phase is plotted as a function of frequency for different values of differential TEC (dTEC; the difference in TEC between antennas, rather than the absolute TEC value), for different bandwidths. Sufficient frequency and time resolution is required to track dTEC and thus the ionosphere. This must be done separately for each isoplanatic patch (a region of constant phase) in the sky, at a time resolution which tracks the changes in dTEC. For Dutch LOFAR this is unproblematic, but the phase variation is faster with time over a smaller region of sky for the international baselines. In practice, we solve for the ionosphere corrections with an iterative approach, solving for the initial bulk correction on an LBCS source, which increases the phase coherence and allows direction-dependent residual corrections using longer solution intervals. 

\paragraph{The Low Band Antenna (LBA)}
Currently, most progress has been made using LOFAR HBA, which operates at frequencies between 120--168~MHz.
However, it is possible to extend this to even lower frequencies, using LOFAR LBA antennas, which operate between 10--80~MHz \citep[e.g.,][]{morabito_lofar_2016,groeneveld_pushing_2022}.
This provides us with unprecedented sharpness at these frequencies, although still a factor $\sim$ 3 lower resolution than with the HBA.
At these frequencies, the challenges of calibration are significantly worse compared to calibration at higher frequencies.
At such low frequencies, the ionosphere becomes especially challenging to calibrate, becoming completely opaque when approaching the ionospheric plasma frequency ($\sim$ 5~MHz, depending on space weather). Ionospheric phase offsets scale inversely with frequency, meaning that an effect at HBA frequencies is three times more pronounced at LBA frequencies.
The ionosphere sets the isoplanatic patch size for LOFAR, and this can decrease rapidly with decreasing frequency and/or poor ionospheric conditions, resulting in less than one usable calibration source per patch in some cases. 
So far, this effect has significantly limited our ability to produce sharp radio images with LOFAR LBA, compared to HBA. Additionally, sources often have low-frequency turnovers, which means they become fainter, in turn lowering the signal-to-noise achievable for calibration.

\subsection{Methods}
\label{subsec:methods}
The general approach for high-resolution imaging with the HBA of the ILT is to build off the calibration used for the more `standard' operations using only the Dutch stations, e.g. for LoTSS. The observing strategy is the same in both cases: typically an 8-hour observation of a field, which is bookended by 10-min scans on standard flux calibrators. We observe two calibrators for redundancy, in case there is a failure during one of the observations, or if the data cannot be properly calibrated. The standard flux calibrator is processed as described in \cite{de_gasperin_systematic_2019}, including the international stations. High-resolution sky models are incorporated in the LOFAR Initial Calibration (LINC) pipeline\footnote{\href{https://git.astron.nl/RD/LINC}{https://git.astron.nl/RD/LINC}} for the following sources: 3C 295, 3C 196, and 3C 48. These sources are bright enough to dominate the calibration, and are bright enough to provide sufficient signal-to-noise for the LINC pipeline. 

After solving for instrumental effects (bandpass, polarisation alignment, and clock offsets) on the flux calibrator, these solutions are transferred to the data for the target field. From there, LINC is used to find direction-independent phase solutions for the Dutch stations, followed by direction-dependent solutions using the ddf-pipeline \citep[described in ][]{shimwell_lofar_2019,tasse_lofar_2021}. Following this, the LOFAR-VLBI pipeline\footnote{\href{https://git.astron.nl/RD/VLBI-cwl}{https://git.astron.nl/RD/VLBI-cwl}} initially prepares the data by applying all solutions from the calibrator (for all stations) and the target field (for Dutch stations), selecting in-field calibrator candidates from LBCS, and then performing initial corrections for the international stations on the best in-field calibrator \citep[see][for more details]{morabito_sub-arcsecond_2022}. These sources must be bright enough to provide initial solutions for the field. 

Once the bulk direction-independent corrections have been applied, imaging of small fields of view for sources in the field can be accomplished. Self-calibration of these individual directions (and the in-field calibrator) uses the script developed by \cite{van_weeren_lofar_2021}, and uses bandwidth and time smearing to suppress other sources in the field. Averaging the data, or combining the phase-corrected core stations together into a single station, can help with data volume reduction and increasing signal to noise on individual baselines. Full field of view imaging builds on top of all previous calibrations; \cite{sweijen_deep_2022} demonstrated full field of view imaging at sub-arcsecond resolution, and recently \cite{de_jong_into_2024} combined observations to make the currently deepest wide-field image. \cite{de_jong_into_2024} also demonstrated intermediate resolution (0.\sarc\ 6 and 1.\sarc\ 2) imaging with the same dataset.

\section{What makes the ILT unique even in the next generation}
\label{sec:unique}

Why is sub-arcsecond imaging at 144 MHz worth the effort? Higher frequency instruments can achieve similar or better resolution, and next-generation instruments like the Square Kilometre Array Observatory (SKAO) will achieve better sensitivity in this frequency band. It is the \textit{combination} of MHz frequencies, flexible resolution, and $\sim6\,$degree$^2$ field of view that make the ILT truly transformational. It represents a significant paradigm shift from targeted studies of small samples to blind surveys with large enough numbers for statistical studies. Sub-arcsecond resolution also easily beats the confusion limit compared to the SKAO-Low at the same frequencies \citep[see, e.g., Shimwell et al. submitted,][]{sabater_lofar_2021}. The ILT also provides simultaneous measurements at the same frequency from sub-arcsec up to degree angular scales. It has matched resolution with instruments at higher frequencies and in other wavebands. No other current or planned radio instrument is capable of this. The following sub-sections highlight the unique capabilities of sub-arcsecond imaging with the ILT, using examples from the literature to demonstrate each point. 

\subsection{Multi-wavelength studies: matched resolution with optical, infrared, and X-ray}
\label{subsec:multiwav}
\begin{figure}[h]
    \centering
    \includegraphics[width=0.51\textwidth]{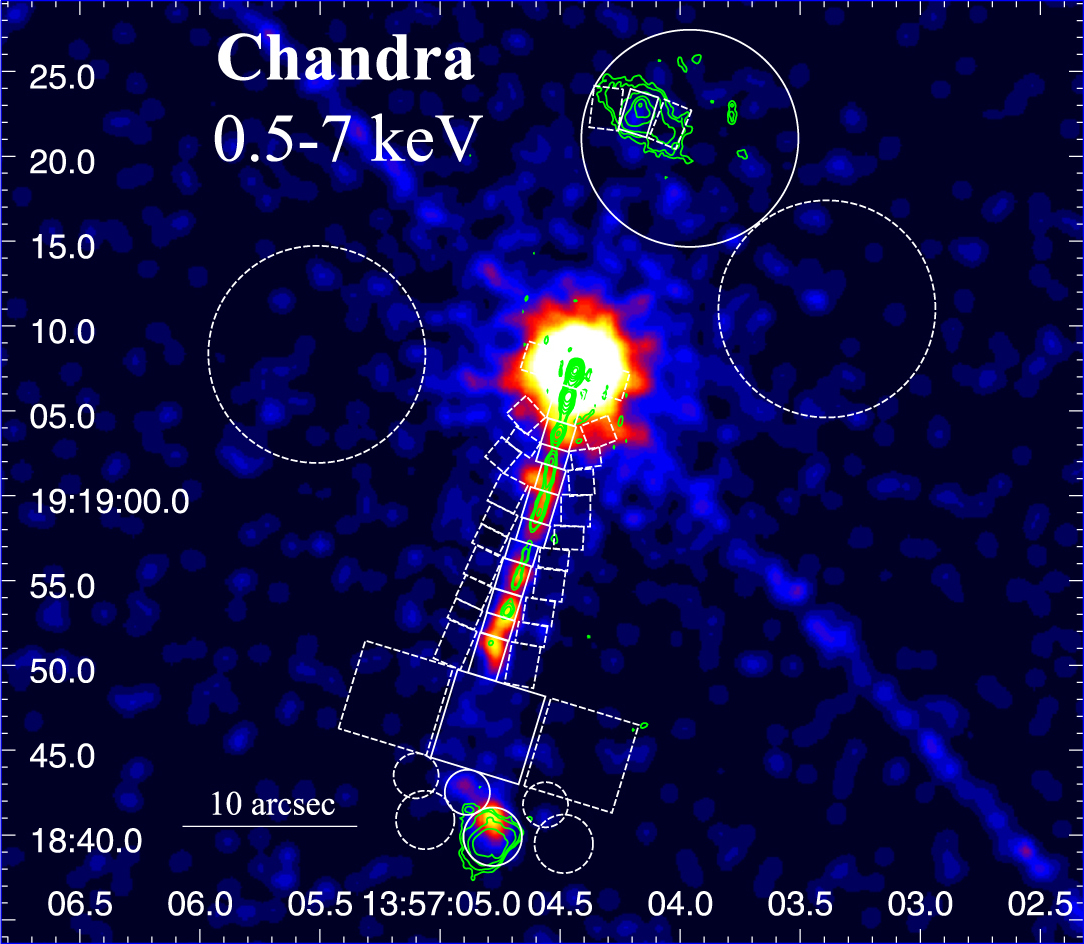} 
    \hfill
    \includegraphics[width=0.47\textwidth]{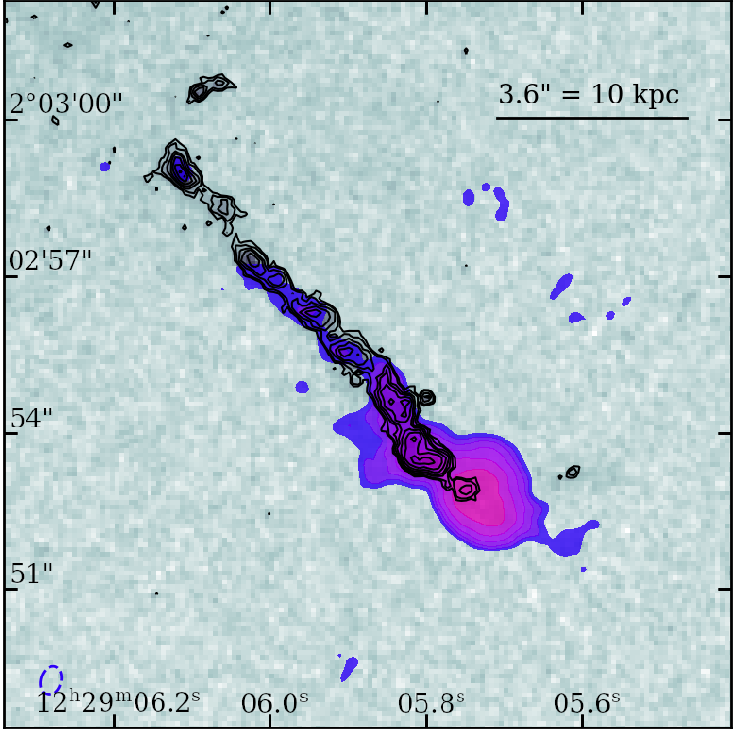} 
    \caption{\textit{Left}: $144\ \mathrm{MHz}$ International LOFAR Telescope (ILT) observation of 4C 19.44 overlaid in green contours on a Chandra observations of the $0.5$-$7\ \mathrm{keV}$ X-ray emission \citep[adapted from][]{harris_lofar_2019}, reproduced by permission of the AAS. \textit{Right}: $144\ \mathrm{MHz}$ ILT observation of 3C 273 overlaid in black contours on Hubble Space Telescope (HST) observations in the $850$–$1700\ \mathrm{nm}$ range (Credit: Harwood et al., A\&A, 658, A8, 2022, reproduced with permission \textcopyright\  ESO.)} \label{fig:lofar_multiwav}
\end{figure}

At sub-arcsecond angular resolution, the ILT is well-matched with the capabilities of optical, infrared, and X-ray instruments. The typical $\sim$0.\sarc\ 3 resolution of the HBA provides Hubble Space Telescope (HST)-like resolution from the ground, outperforming ground-based optical telescopes in sub-optimal seeing conditions. This enables multi-wavelength studies across decades of frequency or wavelength at similar angular resolutions. Figure~\ref{fig:lofar_multiwav} show two examples. \cite{harris_lofar_2019} combined $144\ \mathrm{MHz}$ HBA data with X-ray data from Chandra to study the jet knots of the quasar 4C 19.44, finding low-frequency turnovers in the knots. Sources like these may help understand the link between radio and X-ray emission via Inverse Compton scattering. \cite{harwood_resolved_2022} combined $144\ \mathrm{MHz}$ HBA data with HST near-infrared observations to study the jet of 3C 273, also finding that the low-frequency spectra showed turnovers due to free-free absorption or synchrotron self-absorption. 

\begin{figure}[h]
    \centering
    	\includegraphics[width=0.7\textwidth,clip,trim=0cm 0cm 11cm 0cm]{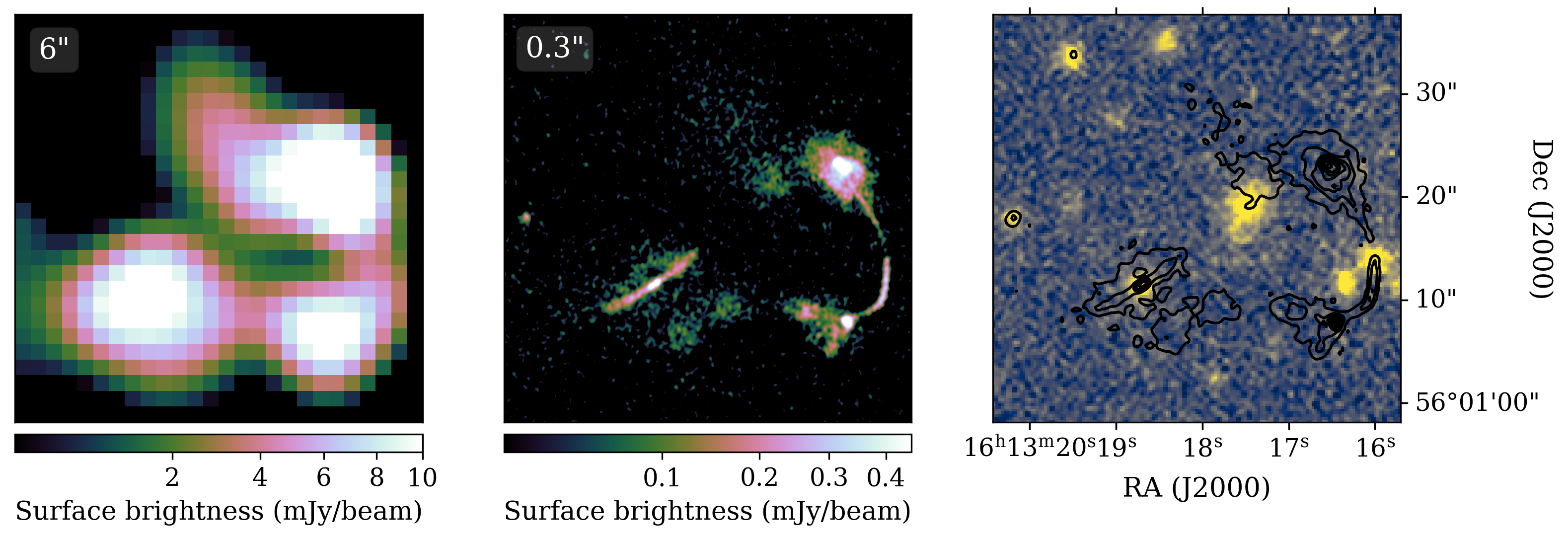}
        \includegraphics[width=0.7\textwidth]{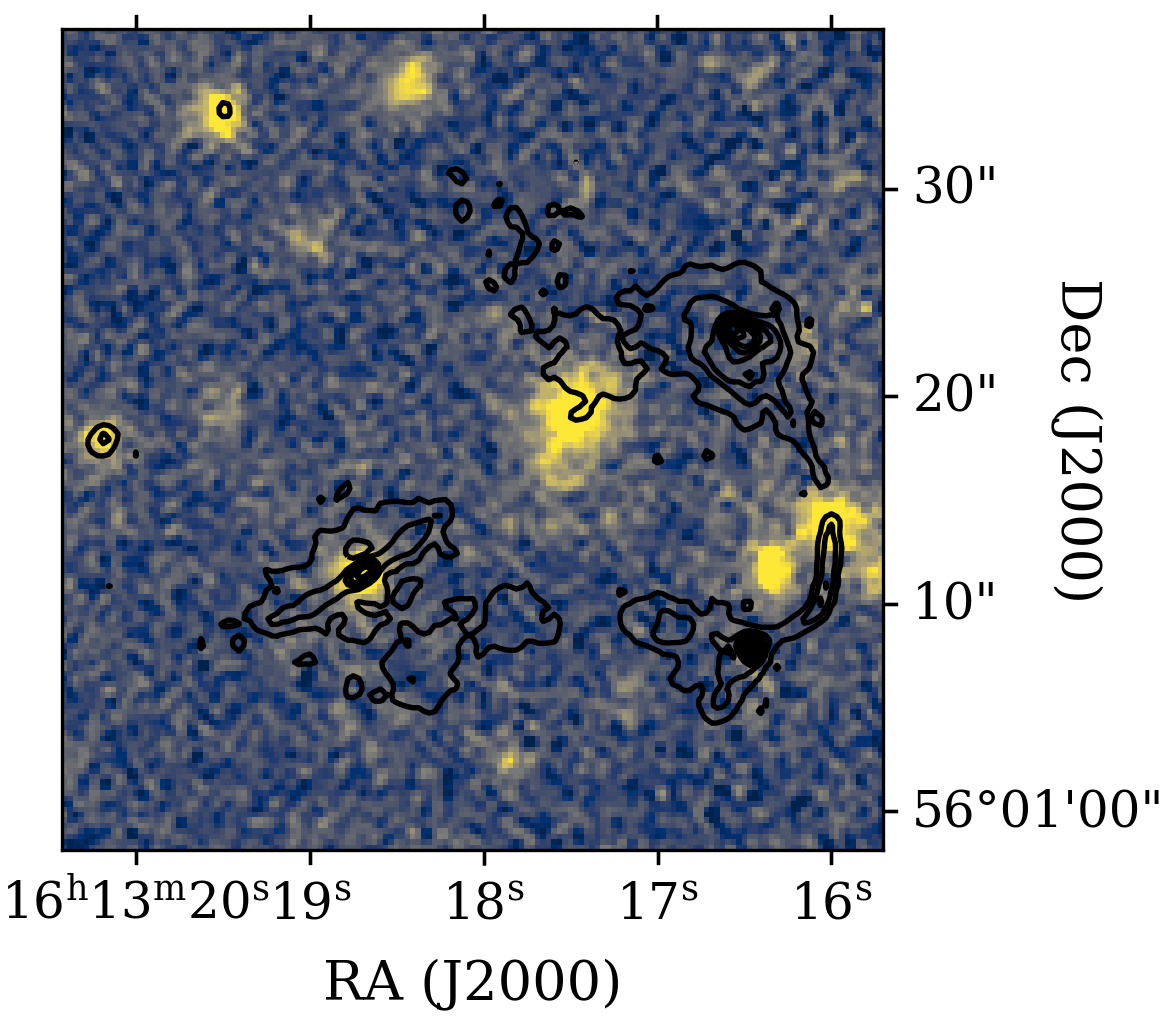}
    \caption{ Sub-arcsecond resolution helps de-blend sources for correct identification of multi-wavelength sources. This also drastically changes the confusion limit, allowing the ILT to still observe individual sources well below the confusion limit of the SKAO-Low. \textit{Top left:} 6\sarc\ \ resolution radio image from \citep{2025arXiv250104093S}, \textit{top right:} 0.\sarc\ 3 resolution radio image from \cite{de_jong_into_2024}, and \textit{bottom: }optical image taken from the DESI legacy survey \citep{dey_overview_2019} with black radio contours from LOFAR data at 0.\sarc\ 6 resolution. }
    \label{fig:lofar_multiwav2}
\end{figure}

This kind of resolution also helps with associating radio emission with the correct multi-wavelength counterpart. In Figure~\ref{fig:lofar_multiwav2}, what appears to be a single wide-angle tail radio source at 6\sarc\ \ resolves into three separate sources: a point source and two Fanaroff-Riley \citep{fanaroff_morphology_1974} type sources. The high resolution image, once properly tied to an astrometric reference frame, allows for proper association with multi-wavelength counterparts. For many projects, accurate registration of ILT images with images at other wavelengths is crucial. Like other radio interferometers, The ILT is in principle capable of astrometry at an accuracy much better than the resolution; however, it is in practice limited by time-dependent position shifts induced by time-dependent ionospheric phase variations. In practice, astrometry is typically good to better than 100\,mas \citep{de_jong_into_2024} and it may be possible to achieve a factor of 2 better than this under good conditions. 

\subsection{A paradigm shift: blind statistical galaxy studies}
\label{subsec:tb}
In radio-quiet AGN, there are several physical mechanisms which could produce synchrotron radio emission \citep[see][for details]{panessa_origin_2019}. Star formation and AGN activity, either from small-scale jets or shocks from AGN-driven winds\footnote{Another possibility is coronal emission but this is thought to be negligible in the MHz frequency regime and we do not consider it here.}, all produce synchrotron radiation and it is difficult to distinguish between them. 
Using only radio luminosity, one can identify AGN as outliers from the radio -- star formation rate (SFR) relation where the radio is far more powerful than what is expected from the SFR derived by another means \citep[e.g.,][]{delvecchio_vla-cosmos_2017,best_lofar_2023}, but finding fainter radio emission due to AGN activity is a challenge. The only way to unambiguously identify radio emission due to AGN in more typical sources is via brightness temperature measurements.  
 Star formation has an upper limit to the surface brightness (flux density per solid angle) it can produce \citep{condon_radio_1992}, and anything above this must be due to AGN activity. The measured surface brightness is typically characterised as a non-physical brightness temperature, $T_b$. This has historically been measured at GHz frequencies with VLBI instruments like the European VLBI Network (EVN), as $T_b \propto \nu^{-1}\,\Omega^{-1}$, where $\nu$ is the observing frequency and $\Omega$ is the beam solid angle. This drives the need for milli-arcsec resolution at GHz frequencies to reach the required sensitivity to compact emission, as characterised by $T_b$. 
 
\begin{figure}
\includegraphics[width=0.65\textwidth,clip,trim=0cm 0cm 0.5cm 0cm]{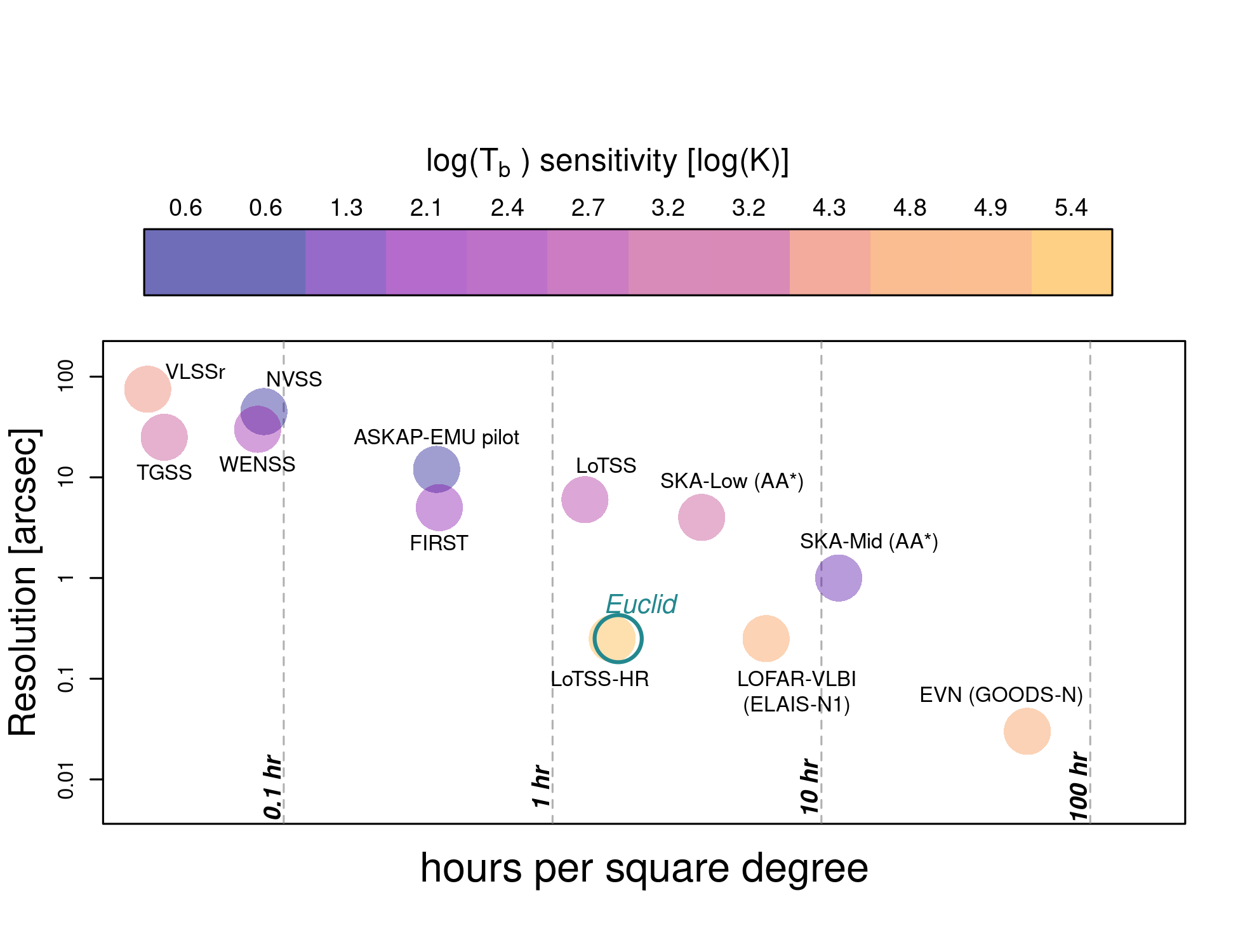}
\includegraphics[width=0.34\textwidth]{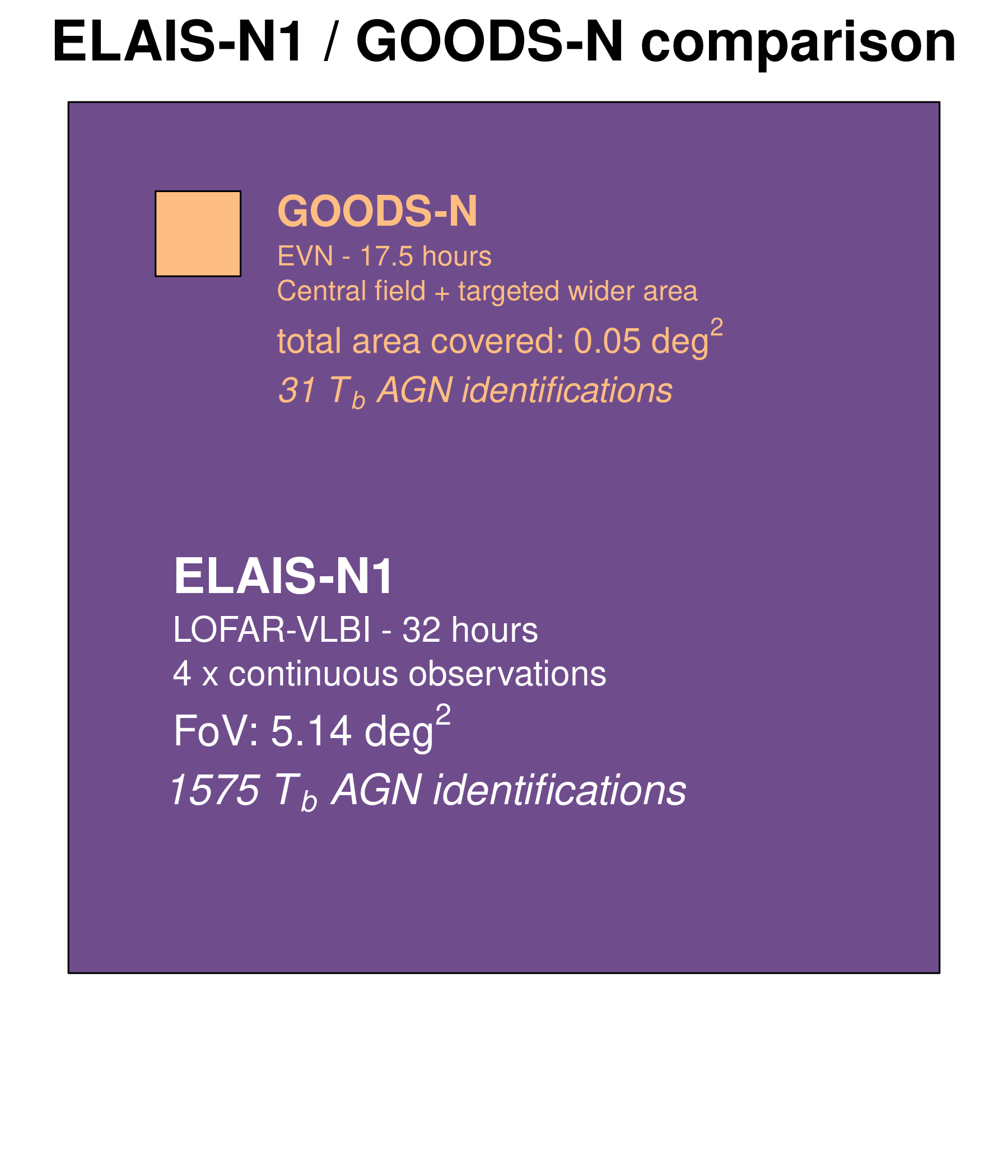}
 \caption{\textit{Left:} Comparison of the field of view, resolution, and sensitivity to compact sources (i.e., $T_b$), for a variety of different surveys. The x-axis shows the number of hours it would take to cover one square degree. Surveys further to the left have larger fields of view and require very little time to cover this area, while surveys to the right require multiple observations mosaicked over an area larger than the intrinsic telescope field of view, and require up to hundreds of hours to cover a square degree. The y-axis shows the resolution which can be achieved. The colours of the points correspond to sensitivity to compact structure, converted to brightness temperature using the frequency and resolution of the survey. Euclid is shown for context on the plot as an open green circle. \textit{Right:} A comparison of deep surveys with the LOw Frquency ARray (LOFAR) and the European VLBI Network (EVN), which are the only instruments/surveys able to do active galactic nucleus (AGN) identifications via $T_b$. The surveys are of GOODS-N \citep{radcliffe_nowhere_2018} and ELAIS-N1 \citep{de_jong_into_2024}, and both use $\sim$32 hours of data. The resulting sky coverage is visually indicated by the size of the orange and purple boxes, respectively.  }
 \label{fig:tb}
\end{figure}

At 144 MHz the ILT reaches similar sensitivity to compact emission with the sub-arcsec resolution of the ILT and these high-resolution images can be used for unambiguous AGN identification \citep{morabito_identifying_2022}. \textit{The paradigm shift for $T_b$ identification of AGN at low frequencies is the increased field of view compared with higher observing frequencies. }
Figure~\ref{fig:tb} demonstrates the advantage which the ILT has in terms of resolution, field of view, and sensitivity to compact emission (characterised by $T_b$). The left panel shows resolution vs. number of hours of observation time required to survey 1 degree$^2$, for a number of surveys. The points are coloured by the survey's sensitivity in terms of $T_b$. Euclid is also plotted as an open circle for comparison in terms of resolution and field of view. Only LOFAR high-resolution surveys (LoTSS-HR, ELAIS-N1) and the EVN survey of GOODS-N reach the $T_b$ sensitivity required to identify AGN emission ($\sim 10^5\,$K, but this is frequency dependent). In the right panel is a more direct comparison of the ELAIS-N1 field recently published by \cite{de_jong_into_2024} and the GOODS-N field published by \cite{radcliffe_nowhere_2018}. ELAIS-N1 uses 4 repeat observations with the same observational setup which is 32 hours in total, while the GOODS-N survey covered a central 7.$^{\prime}$5 radius area with targeted observations of bright sources up to 25$^{\prime}$ from the centre, using 17.5 hours. The figure shows an accurate representation of the relative size of each field. The area covered by ELAIS-N1 is 105 times larger than that of GOODS-N, with 51 times more AGN identifications from $T_b$ measurements. 

While the sky density of $T_b$-identified AGN in GOODS-N is higher, the ILT observations of ELAIS-N1 offer, for the first time, a wide-area blind survey which has enough $T_b$ sensitivity for identifying AGN. This shifts us from small samples and/or targeted observations to sample sizes which are large enough for statistical studies of radio emission in the underlying galaxy population. For galaxies where AGN components have been identified via $T_b$ measurements in the high-resolution images, comparison with the low-resolution (6\sarc\ ) images \textit{made from the same observational dataset} provide the means to simultaneously measure the radio emission from both AGN activity and star formation. This is another advantage over GHz VLBI instruments, where information at different scales (see \S~\ref{subsec:dynamicrange}) must be obtained from different instruments, which can introduce systematics. \cite{morabito_hidden_2025} used these capabilities to construct, for the first time, radio luminosity functions constructed by physical process, star formation or AGN activity, showing that there is 50 percent more radio emission than expected from AGN activity. Luminosity functions require large statistical samples, and are not currently possible with any other instrument with sufficient $T_b$ for identifying AGN. 

\subsection{Spectral modelling: the low-frequency spectral anchor}
\label{subsec:specmodel}
Synchrotron radiation initially yields a power law spectrum, but it will not keep this shape forever. Particle acceleration and energy loss processes can modify the shape of the observed spectrum. At low frequencies, the spectrum is changed by a low energy cut-off in the electron population, or absorption in the form of free-free absorption or synchrotron self-Compton scattering. At high frequencies, the power law can become curved due to higher-energy electrons radiating away their energy faster than lower-energy electrons, which is referred to as ``spectral ageing''. Spectral modelling is the practice of modelling these processes in order to fit the observed spectrum and extract physical properties of the radio-emitting plasma. Physically meaningful modelling is only possible if one has spatially resolved information \citep{harwood_spectral_2013,harwood_spectral_2015,harwood_spectral_2017} across a range of frequencies; the low frequencies (below $\sim$ 400 - 500 MHz) are absolutely crucial to help fix the injection index (the initial spectral index before any ageing has happened). The ILT is the only instrument capable of providing this spatially resolved information at these frequencies for anything but the most nearby sources, which has been used in several studies \citep{sweijen_high-resolution_2022,harwood_resolved_2022,bonnassieux_spectral_2022, cordun_vlbi_2023}.

\begin{figure}
    \centering
    \includegraphics[width=0.8\textwidth]{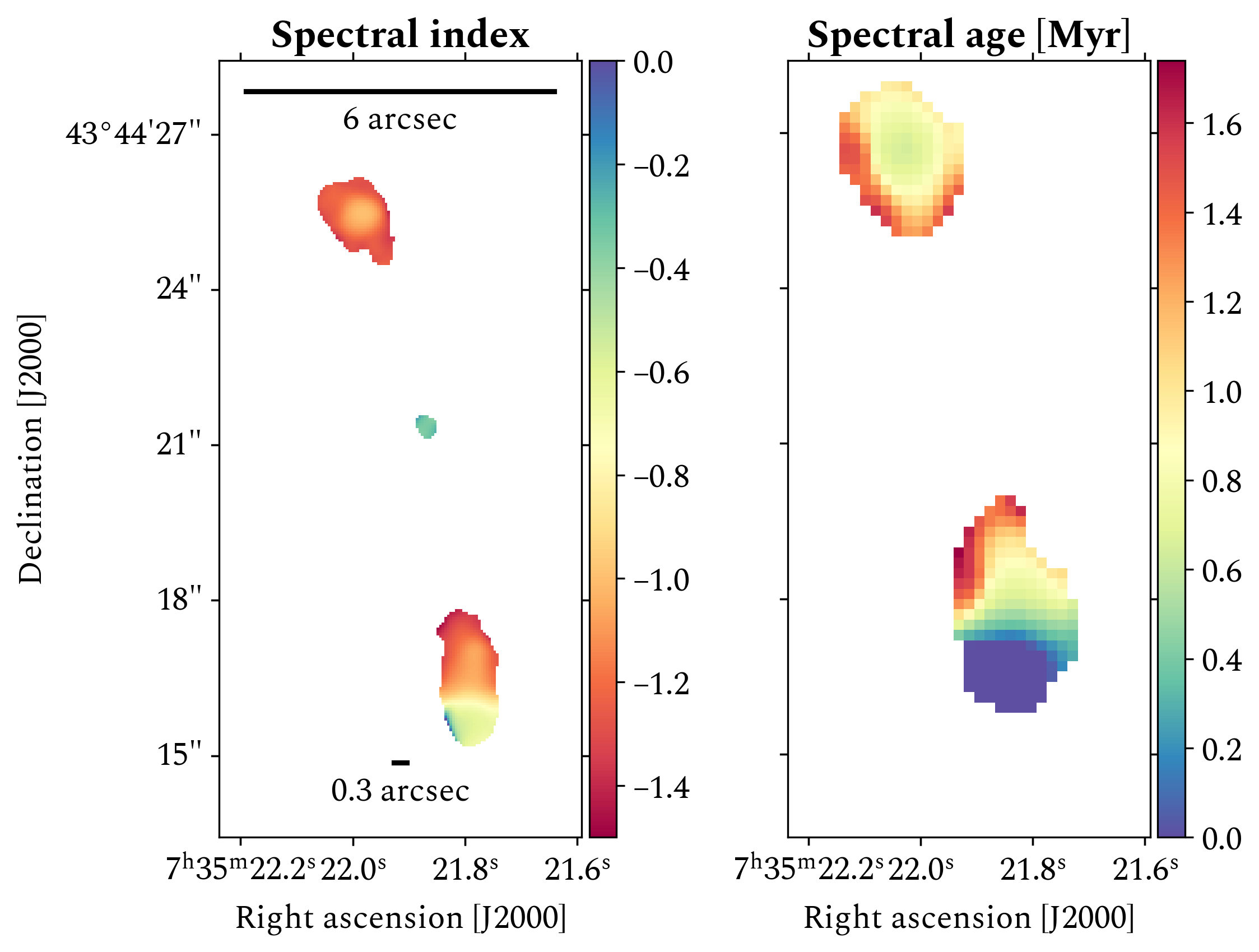}    
    \includegraphics[width=0.75\textwidth]{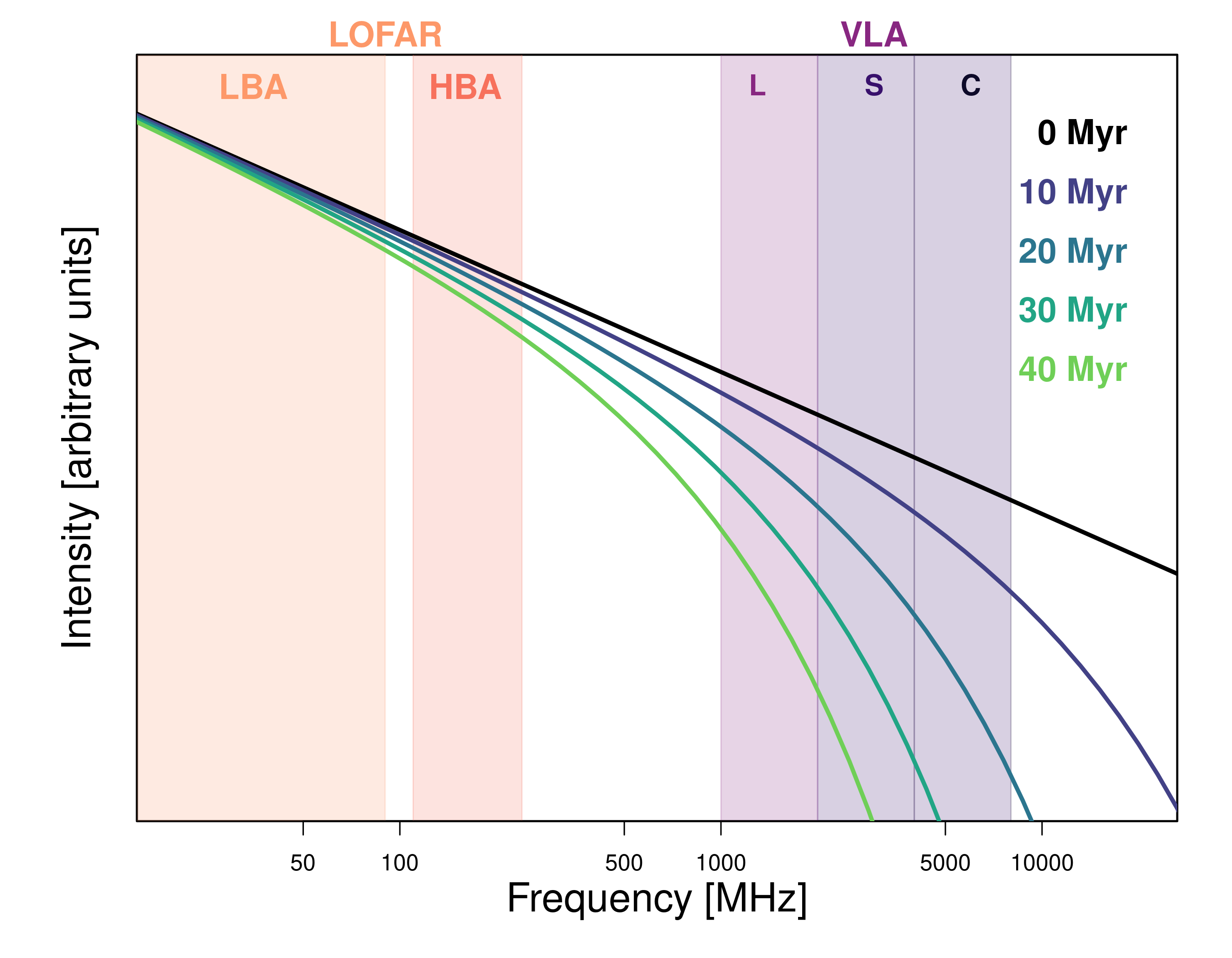}
    \caption{Spatially resolved spectral indices and ages of 4C 43.15 at $z = 2.4$ adapted from \cite{sweijen_high-resolution_2022} \textit{Top left}: spectral indices at a resolution of $0.38''$ from a least-squares fit between the $144\ \mathrm{MHz}$ HBA and $4.7$ and $8.4\ \mathrm{GHz}$ Very Large Array (VLA) data. Bluer and redder colours reflect flatter and steeper spectral indices, respectively. \textit{Top right}: spectral ages at a resolution of $0.9''$ derived from fitting a JP-Tribble model to the $54$ and $144\ \mathrm{MHz}$ International LOFAR Telescope (ILT) data and $4.7$ and $8.4\ \mathrm{GHz}$ Very Large Array (VLA) data. Bluer and redder colours reflect younger and older spectral ages, respectively. \textit{Bottom:} A JP model with injection index of 0.6, $B$-field of 10$^{-9}\,$T, for a range of spectral ages. The frequency ranges of LOFAR's Low Band Antennas (LBA) and High Band Antennas (HBA) are shown as shaded regions, as are the L, S, and C bands of the VLA, which have near-matched resolution (or can be used to produce near-matched resolution with imaging options like $uv$ cuts). This shows the need for low frequencies to properly constrain the spectral age when modelling. Models produced using BRATS \citep{harwood_spectral_2013,harwood_spectral_2015}. Top panels credit: Sweijen et al., A\&A, 658, A3, 2022, reproduced with permission \textcopyright\  ESO.}
    \label{fig:4c43.15}
\end{figure}

Low-frequency instruments become more and more important as one studies sources at higher and higher redshifts and the rest-frame frequencies probed increase by $1 + z$. This is demonstrated using, for example, the high-redshift source 4C 43.15 located at $z = 2.4$. For the first time, spatially resolved spectral models could be fit to such a high-z object by combining the ILT's LBA data at $54\ \mathrm{MHz}$ (\cite{morabito_lofar_2016} and HBA data at $144\ \mathrm{MHz}$ (\cite{sweijen_high-resolution_2022} with $1.4\ \mathrm{GHz}$, $4.7\ \mathrm{GHz}$ and $8.4\ \mathrm{GHz}$. Figure~\ref{fig:4c43.15} shows the spectral index and spectral age maps obtained using the \textit{Broadband Radio Astronomy ToolS}\footnote{\url{https://www.askanastronomer.co.uk/brats/}} (BRATS) package \citep{harwood_spectral_2013,harwood_spectral_2015}. The spectral index map demonstrates matched resolution between the ILT at $144\ \mathrm{MHz}$ and the VLA at $4.7$ and $8.4\ \mathrm{GHz}$. For the spectral age map, the resolution is limited by the $54\ \mathrm{MHz}$ image which provides essential constraints for the power-law part of the spectrum, but nonetheless demonstrates the ability for spatially resolved spectral ages in distant objects.

\subsection{Rare objects}
\label{subsec:rare}

For a given interferometer array, the number of objects detected in a survey increases linearly with the area of the survey, which is directly proportional to the time it takes to observe. For typical radio source populations, the number of sources increases inversely with the limiting flux density, and hence is proportional to the inverse square root of the observing time. The most successful surveys, in terms of the number of objects, are therefore wide-field ones, as the number of sources increases more rapidly than going deeper. Large numbers of sources are useful for statistical samples (see Section~\ref{subsec:tb}), but also for detecting intrinsically rare objects and increasing the space for serendipitous and unexpected discoveries.

\begin{figure}[h]
    \centering
    \includegraphics[width=0.7\textwidth]{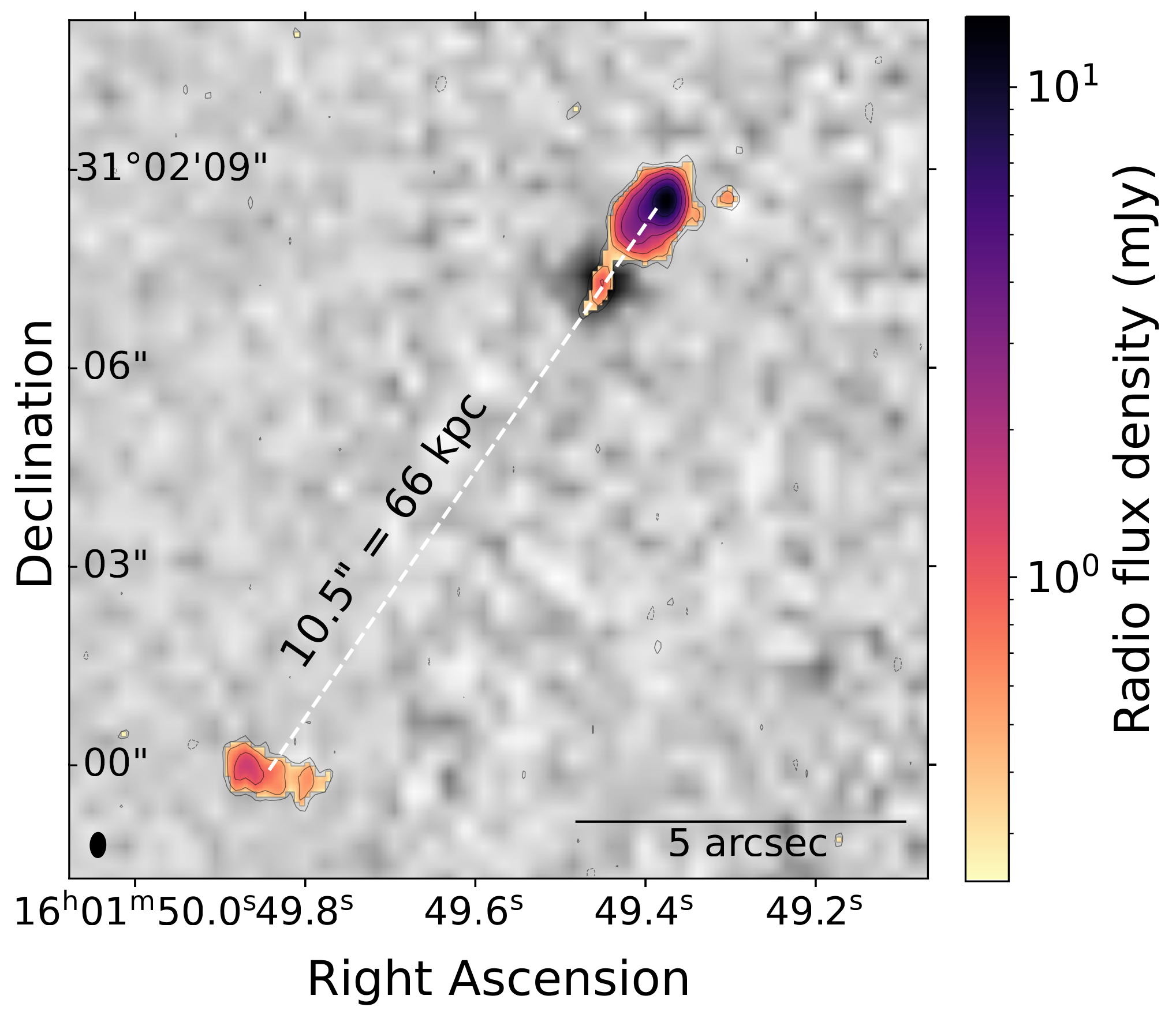} \\
    \hspace{-2cm} \includegraphics[width=0.6\textwidth]{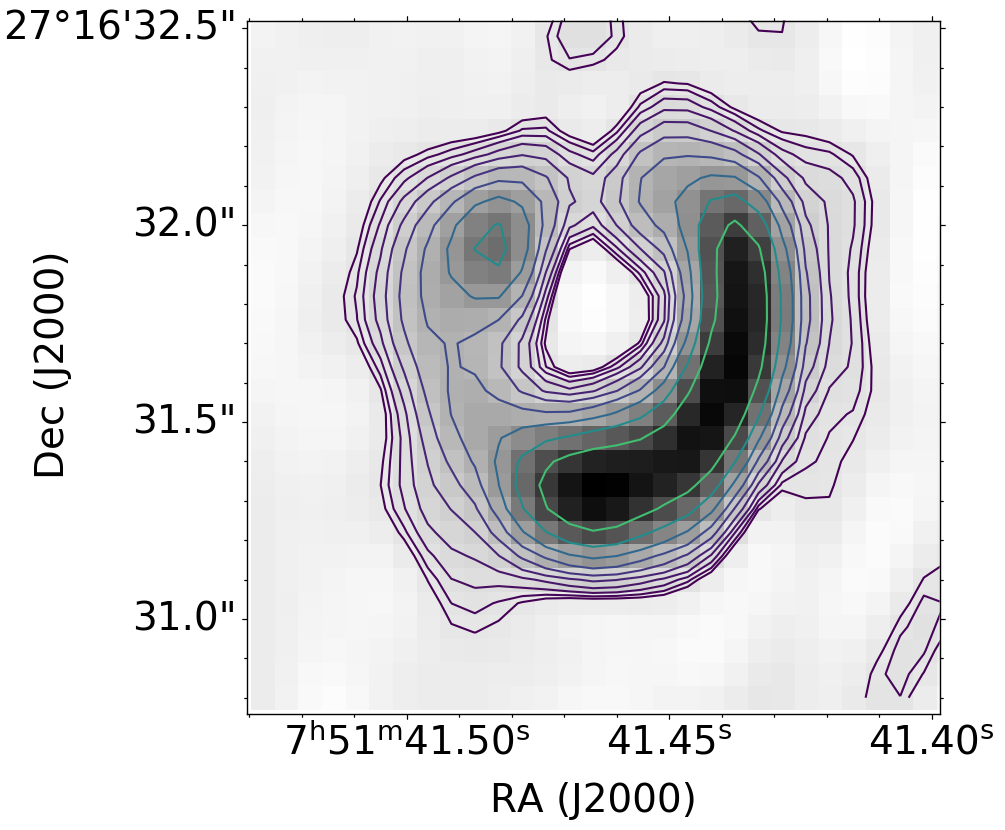}
    \caption{\textit{Top}: the recently discovered radio-loud quasar J1601+3102 at a redshift of $z = 4.912$, adapted from \citep{gloudemans_monster_2024}, reproduced by permission of the AAS. \textit{Bottom}: LOFAR ILT image of the strong gravitational lens system MG\,0751+2716 \citep{badole_high-resolution_2022}}
    \label{fig:rare_objects}
\end{figure}

As one of the instruments capable of wide fields of view, LOFAR is a forefront survey instrument. The ILT's long baselines give it the unique capacity to combine large fields of view with high resolution and astrometric accuracy. This allows not only the discovery of rare objects, but also the immediate characterisation of their morphology and comparison with data at other wavelengths, often at equal or better resolution. One such object is the radio-loud quasar J1601+3102 recently discovered by \cite{gloudemans_discovery_2022}, at a redshift of $z = 4.912$. Originally discovered in LoTSS, high-resolution follow-up with the full ILT \citep{gloudemans_monster_2024} revealed that a neighbouring ``source'' was actually one of the radio lobes, making it the largest known jet in quasars at $z > 4$ to date. The left panel of Fig.~\ref{fig:rare_objects} shows the source, highlighting how the high resolution of the ILT is essential to resolve the southern source and reveal its nature as the southern lobe of the quasar.

Strong gravitational lenses are systems in which a background source is multiply imaged by the gravitational field of a foreground galaxy (the ``lens'') with typical image separations of $0.5-1^{\prime\prime}$. Typically only one in a thousand $z\sim2$ sources lies close enough to a line of sight of a foreground lens to be multiply imaged.  However, they are important for many reasons, including the investigation of mass distributions of lensing galaxies at high enough redshifts that dynamical measurements are difficult, and the magnified views of intrinsically ultra-faint background objects that they provide. In addition, they are cosmologically significant in their ability to provide constraints on $H_0$, if the background object is variable so that time delays between paths to the individual lens images can be determined \citep{refsdal_possibility_1964}, and on $w$ in the rare cases where two sources at different redshifts are seen behind a lens \citep{collett_constraining_2012}.

Some existing lens systems have already been successfully imaged with the ILT, like the one in Figure~\ref{fig:rare_objects} \citep{badole_high-resolution_2022}. Pointed, ILT-resolution observations, together with a lens model, were used to uniquely probe the structure of the background radio source at very low radio frequencies. Low-frequency, high-resolution observations can also be used to study propagation effects such as scattering in the ionized component of the lensing galaxy, since such effects are strongest at low radio frequency and manifest themselves by differences in images originating in the same source but which have followed different light paths through the lens \citep{biggs_radio_2004,badole_high-resolution_2022}. Further work with wide-field ILT observations will result in the discovery of many more lenses; radio flux has already been detected from some optical lens systems with LoTSS-6$^{\prime\prime}$\citep{mckean_gravitational_2021} at the level of a few tens of $\mu$Jy. Since about 300 sources per square degree are detected in standard LoTSS observations \citep{sweijen_deep_2022}, we expect hundreds of strong lens systems in the full LOFAR-ILT survey area. The high ILT resolution then allows characterisation of the nature of the radio emission in sources which, because of the lens magnification, will only otherwise be accessible to the full SKA; this will allow us to probe the relative presence of star-formation \citep{badole_vla_2020} and AGN \citep{hartley_strong_2019} within the faintest high-redshift radio sources.

\subsection{Dynamic range of spatial scales}
\label{subsec:dynamicrange}

The ILT features a unique wide range of baseline scales, providing it with sensitivity to both small-scale and large-scale emission simultaneously. Especially at low radio frequencies, this capability is very valuable as the low-frequency sky contains both small-scale structure \citep[e.g., AGN jets;][]{mingo_revisiting_2019,mingo_accretion_2022} as well as large-scale structure \citep[e.g., radio halos and the discovery of radio bridges;][]{van_weeren_diffuse_2019,govoni_radio_2019}, sometimes even in the same source \citep{kukreti_unmasking_2022}. 
Leveraging the ability to detect a wide range of angular scales with a single observation has recently enabled several breakthrough publications.

\begin{figure}
    \centering
    \includegraphics[width=\linewidth]{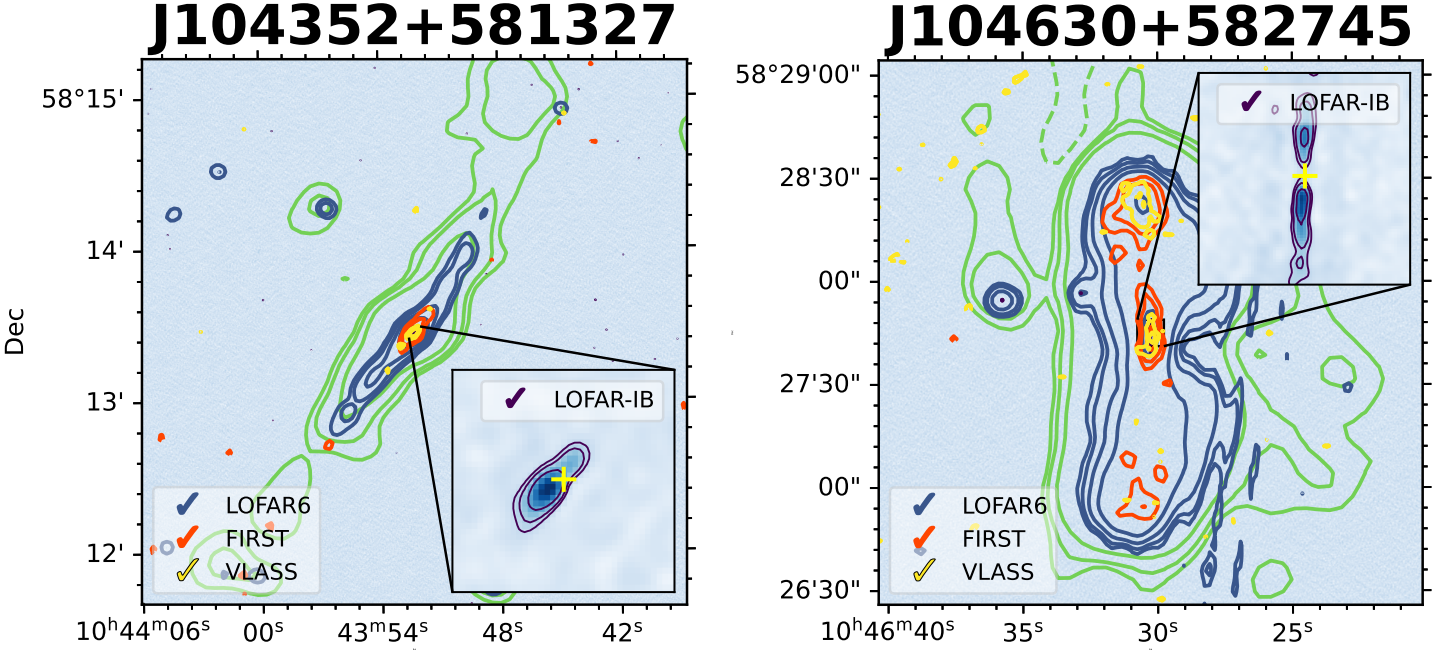}
    \caption{Radio images of active radio galaxies taken with the International LOFAR Telescope (ILT). The green contours show the 144~MHz emission as detected by LOFAR with an angular resolution of 18". The blue contours show the 144~MHz emission as detected by LOFAR with an angular resolution of 6". The red and yellow contours show the radio emission detected by the Faint Images of the Radio Sky at Twenty-cm (FIRST) survey and the Very Large Array Sky Survey (VLASS), respectively. The insert shows the LOFAR image taken with international baselines included.}
    \label{fig:jurlin}
\end{figure}

For studying AGN jets, \citet{de_jong_cosmic_2024} showed that relying only on short Dutch baselines can introduce selection biases due to resolution limits when examining the evolution of jet morphologies up to redshift $z=2.5$. In contrast, recent work by \cite{jurlin_nuclear_2024} illustrates the advantages of combining both short Dutch and long international baselines. ILT observations were in this study able to simultaneously detect recent activity of AGNs close to their AGN cores using the long baselines as well as old emission from previous outbursts using the short baselines. Two examples of such detections are shown in Figure~\ref{fig:jurlin}. The wide range of baselines ensured that both the new activity, as well as the old emission, could be fully detected, and sufficient freedom was available to image at several angular resolutions.

\begin{figure*}
    \centering
    \includegraphics[width=0.48\textwidth]{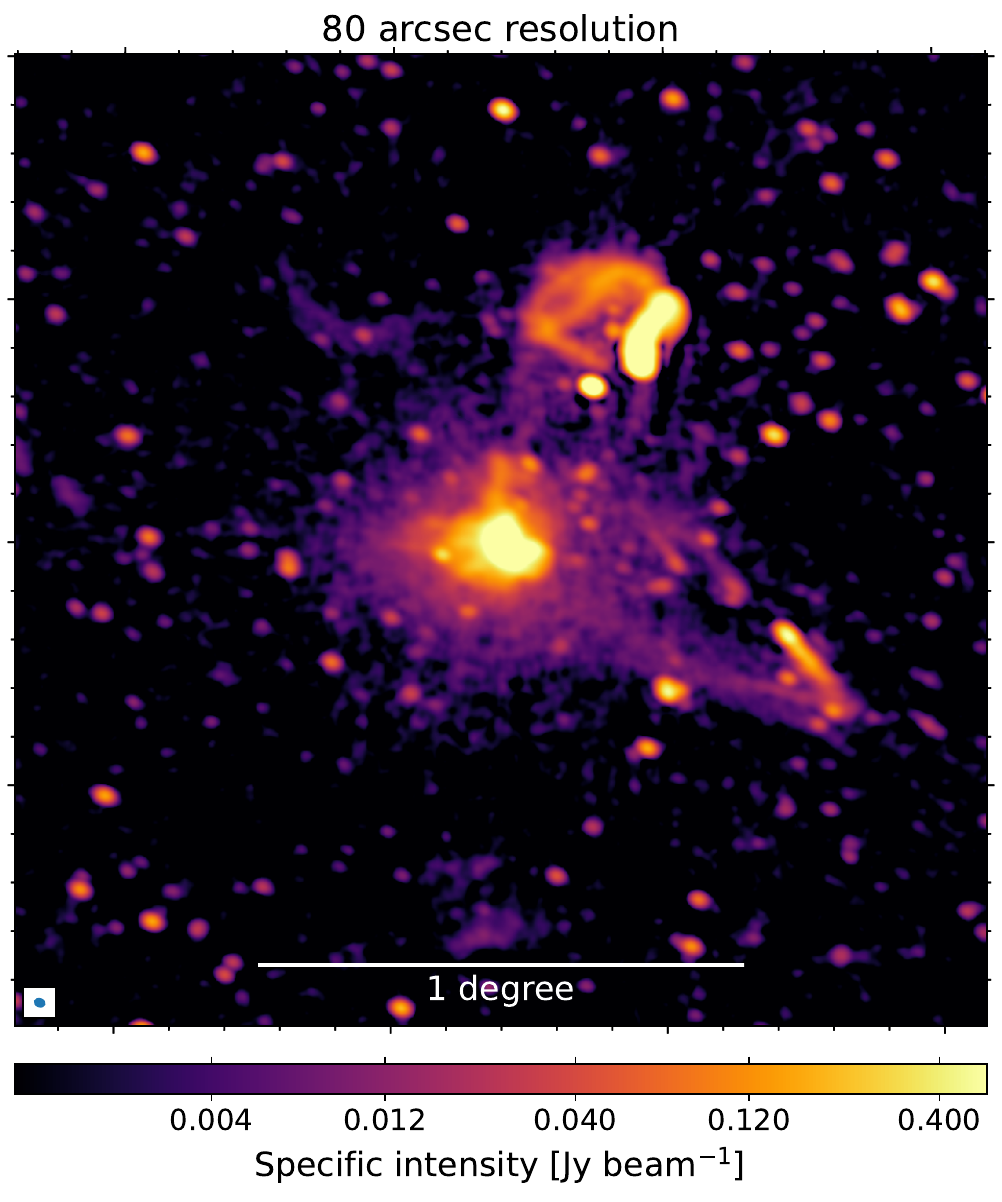}
    \includegraphics[width=0.48\textwidth]{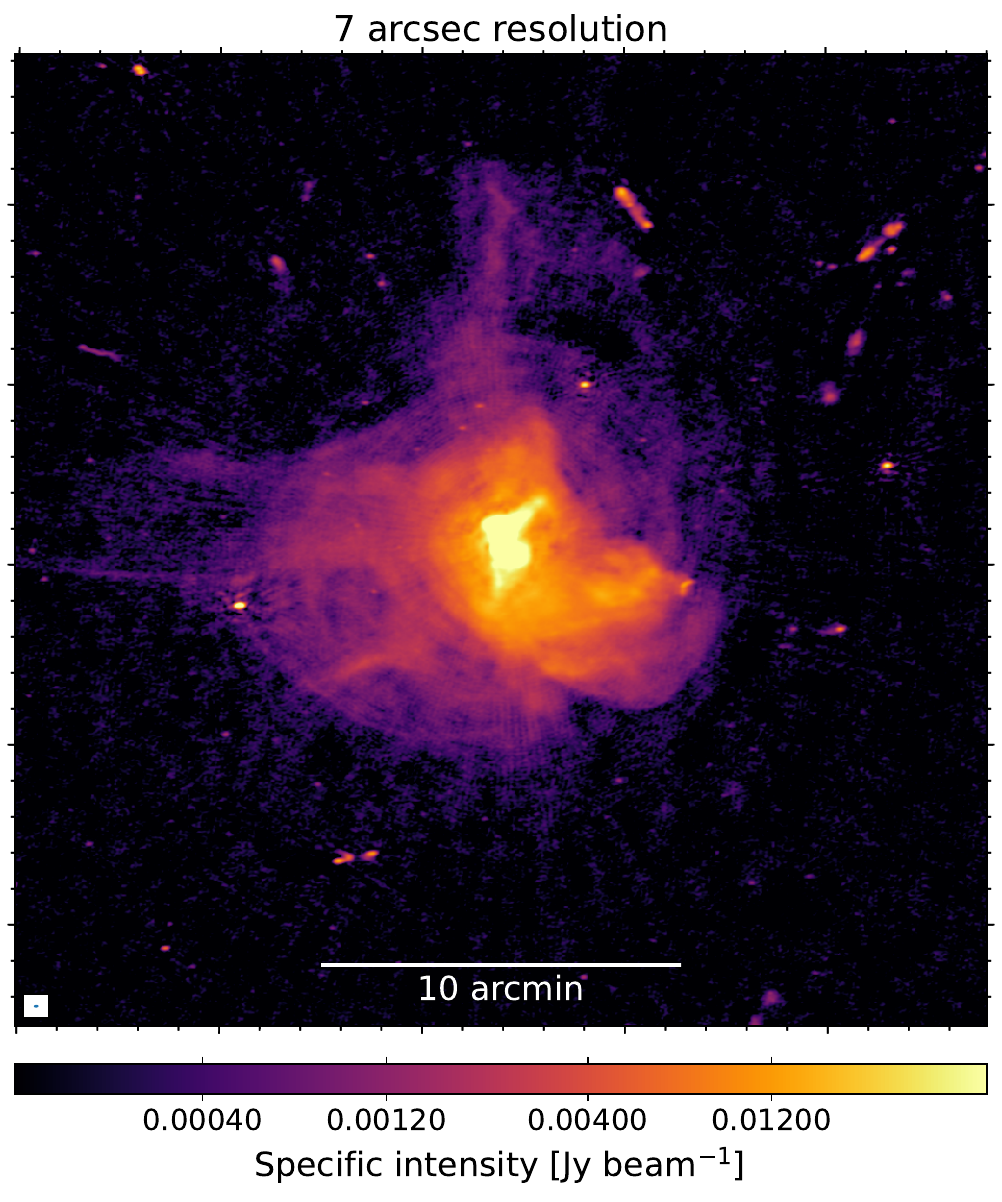}
    \includegraphics[width=0.48\textwidth]{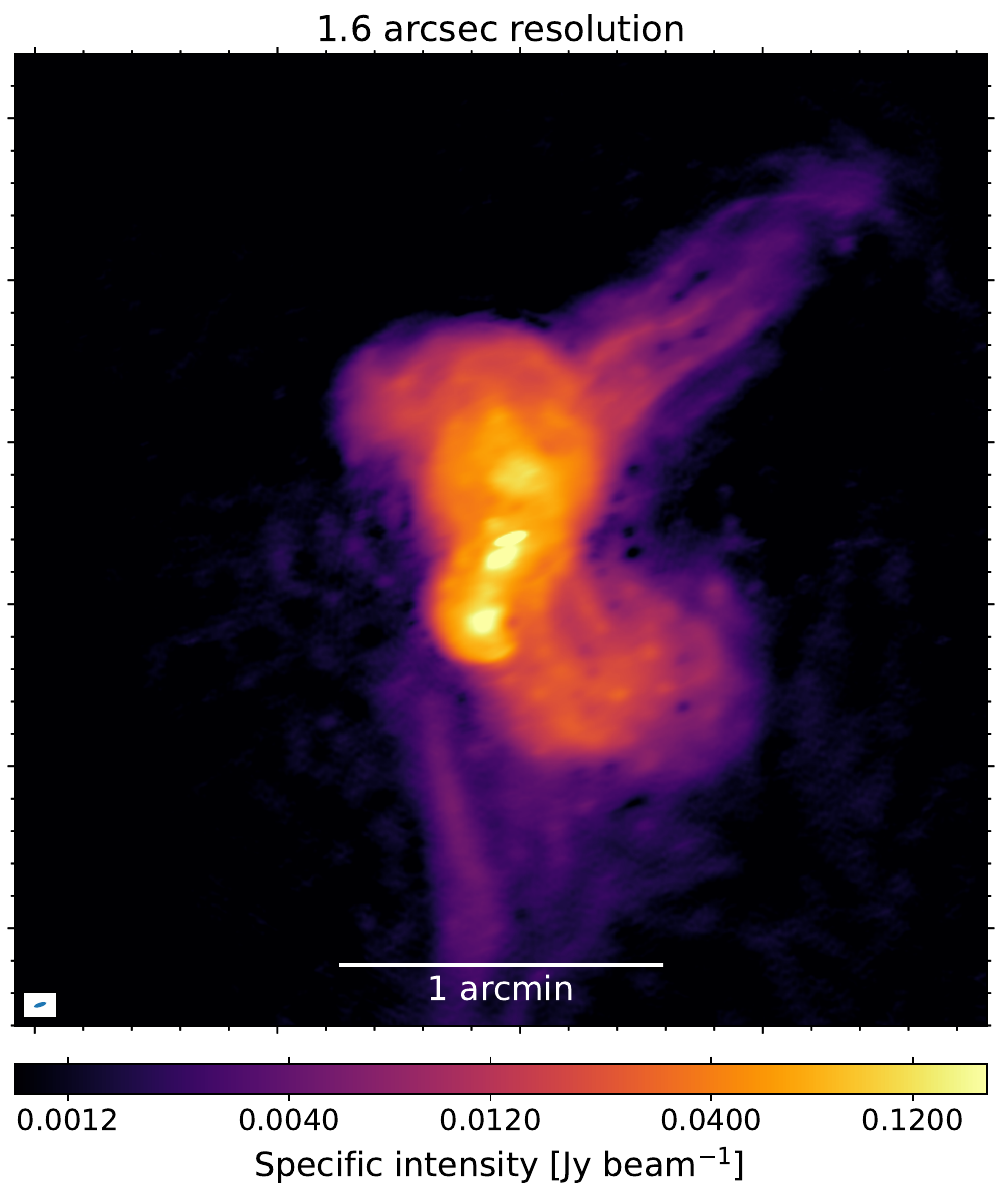}
    \includegraphics[width=0.48\textwidth]{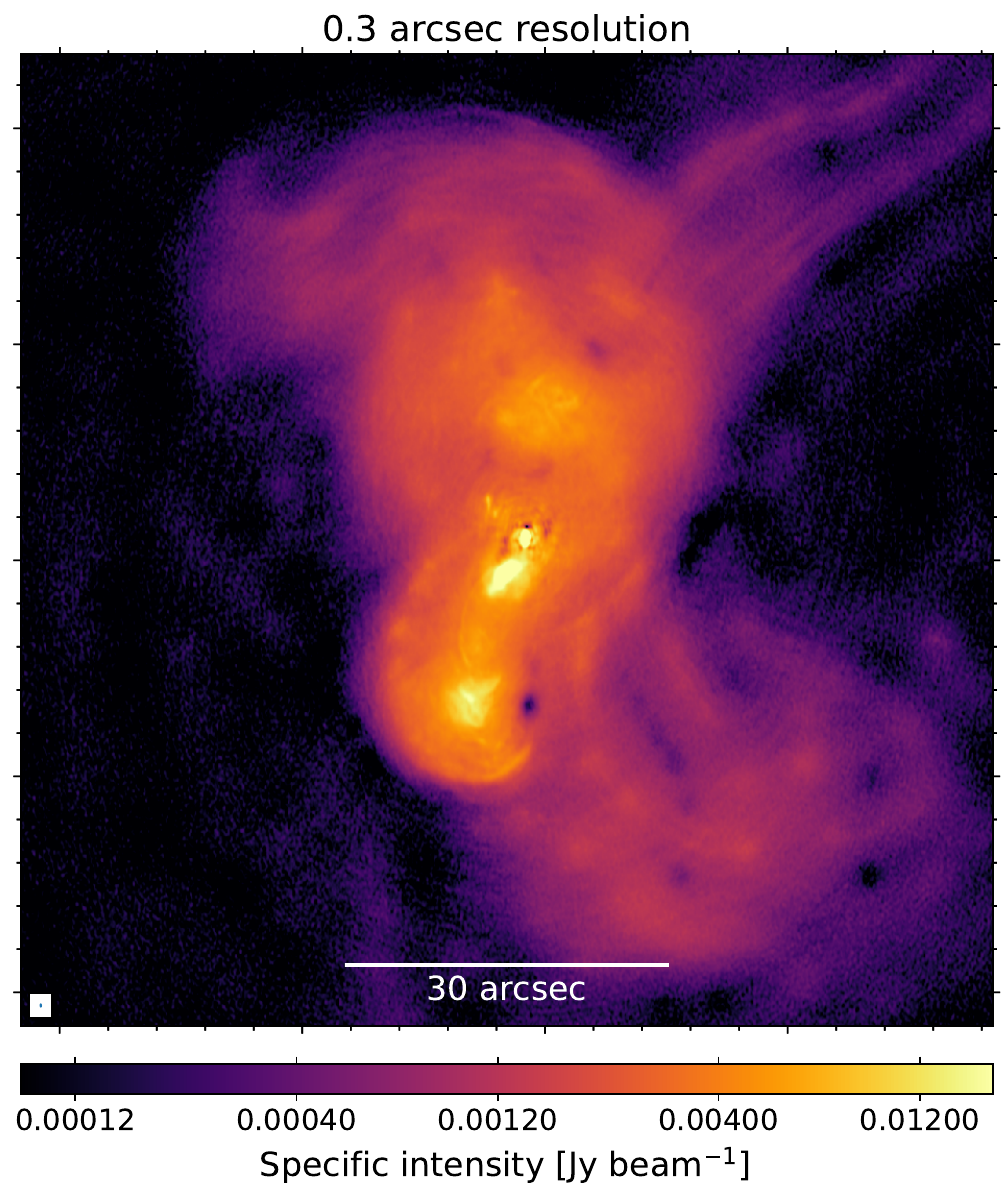}
    \caption{Images of the Perseus cluster demonstrating the exceptional range of resolutions provided by the ILT from a single observation. At a very low resolution of $80^{\prime\prime}$ (\textit{top left}) extended diffuse emission is detected associated with the hot intracluster medium. Several tailed radio galaxies can also be seen.  A higher resolution of $7^{\prime\prime}$ (\textit{top right}) reveals intricate details of the central diffuse radio mini-halo. Additionally, the AGN 3C\,84, associated with the brightest cluster galaxy NGC\,1275, produces bright compact radio lobes at the cluster center. These top-panel images are derived using only LOFAR's Dutch baselines. The bottom-panel images, incorporating the longer European baselines, reveal even finer details, including filamentary structures in 3C\,84, at resolutions of $1.6^{\prime\prime}$ (\textit{bottom left}) and $0.3^{\prime\prime}$ (\textit{bottom right}). More details about the processing and images can be found in \cite{van_weeren_lofar_2024}. }
    \label{fig:perseus}
\end{figure*}

Another prominent example of combining multiple resolutions is the Perseus cluster, as imaged by \citet{van_weeren_lofar_2024}, and shown in Figure~\ref{fig:perseus}. This cluster has complicated previous work due to the presence of structure on a wide range of angular scales. The central radio source 3C\,84, associated with the brightest cluster galaxy (BCG) NGC\,1275, consists of a bright unresolved AGN core in between its bright kpc-scale radio jets and lobes. Just beyond these central structures, old diffuse emission from previous outbursts of the AGN is expanding within a mini halo. Towards the outskirts of the cluster core, a Mpc-scale radio halo can be found. It was the combination of sensitivity on all relevant scales that enabled the discovery of the Mpc-scale radio halo, which previously remained hidden beside the bright compact core. A study on the galaxy cluster Abell 2255, employs techniques from wide-field imaging to create an intermediate wide-field image of 1.5 by 1.5 degrees (De Rubeis et al. in prep.). Similarly to the work on the Perseus cluster, this work leverages multiple resolutions to reveal unprecedented insights into emissions at various scales within galaxy clusters, capturing structures from thin filaments trailing radio galaxies with long baselines to diffuse threads extending hundreds of kpc within the intracluster medium with shorter baselines.

\section{Current state and ongoing work}
\label{sec:ongoing}

After 12 years of science operations, LOFAR has been closed to undergo upgrades. At the time of writing, no new data are coming in, but LOFAR2.0 is expected to start operations in 2025. The upgrades mainly focus on the backend electronics, increasing the intake rate for data at the station level. This enables an increased bandwidth, which will be transformational for the LBA. 

While LOFAR2.0 will provide an improved telescope, there are still tens of thousands of observations in the LOFAR Long Term Archive that are appropriate for generating sub-arcsecond resolution. Our experience over the last $\sim$decade indicates that calibration of this data became far more robust after Cycle 4, due to a combination of the expansion of the array to Poland, and better station calibration tables for the international stations. In general, we recommend focusing on observations where 11 or more international stations participated, but this of course depends on the resolution desired. The community is not idle while waiting for LOFAR2.0, and there are several areas of active ongoing work to (a) fully exploit current data and (b) prepare for LOFAR2.0 Large Programmes. There are 15 Large Programmes, mostly imaging surveys, and a large fraction require high-resolution imaging with the ILT to achieve their science goals. There are four different areas of active work at the moment, aimed at providing more science-quality data to the community. These are described below. 

\paragraph{Wide-area surveying}
As the development of the LOFAR-VLBI pipeline has reached the threshold where it can reliably calibrate the international stations for most target fields, progress is currently being made with reprocessing the archival LoTSS data using this pipeline. This project, known as LoTSS-HR (LoTSS-High Resolution), aims to publish cutout images of all sufficiently bright (\(>10\) mJy) radio sources at an angular resolution of 0.3 arcseconds as well as full field of view images at an intermediate resolution of approximately 1.2 arcseconds. As the calibration of LOFAR-VLBI data is prohibitively challenging for an average user in terms of both computational resources and quality control, this forms a major breakthrough in the accessibility of high-resolution radio images at low frequencies. An initial public data release is currently in preparation for the H-ATLAS area \citep{hardcastle_lofar/h-atlas:_2016}, with later data releases planned for the rest of the LoTSS sky coverage. The LoTSS-HR project is set to form the highest-resolution sky survey with a comparably wide effective sky coverage by over an order of magnitude at any radio frequency, and therefore will provide immense scientific value across a wide range of topics.\\

\paragraph{WEAVE-LOFAR}
WEAVE (WHT Enhanced Area Velocity Explorer) is a multi-object spectrograph built onto the William Hershel Telescope in La Palma. A future tiered survey to be conducted by this instrument is WEAVE-LOFAR \citep{smith_weave-lofar_2016}. The selection criteria for the targeting of radio sources in LoTSS-HR is the same as the targeting strategy for the final Wide tier of WEAVE-LOFAR. Therefore, for every source with a sub-arcsecond resolution radio image thanks to the LoTSS-HR project, there will also be high-quality spectroscopic measurements available for these sources. This will allow for precise spectroscopic redshifts to be measured for these sources. The combination of LoTSS-HR and WEAVE-LOFAR will enable a broad range of scientific projects that demand both high-resolution radio imaging alongside high-quality spectra supported by statistically robust sample sizes. \\

\paragraph{Wide-field imaging}
Thanks to the low observing frequency coupled with the size of the international stations, the ILT encompasses a large FoV, enabling the production of wide-field images of around 6.6 degree$^2$ \citep{sweijen_deep_2022}. This enables thousands of sources to lie within an image from a single 8 hour observation. There are, however, additional considerations to take into account when performing calibration on a widefield observation. Directional-dependent (DD) effects vary across the FoV, due to inconsistent ionospheric conditions for example. To correct for DD effects such as this, multiple bright calibrator sources are required across the scope of the observation. Around 20 to 30 calibrator sources are desired per widefield image. These are automatically selected by assessing if there is sufficient S/N to calibrate the international stations \citep{de_jong_into_2024}. Using Voronoi tessellation \citep{schwab_global_1983, van_weeren_lofar_2016}, the image is split into facets, each facet containing a single calibrator source. The current strategy for widefield imaging is to calibrate and image each of these facets individually, and then to mosaic these together to create the final full image.   

One of the major challenges of producing sub-arcsecond widefield images comes from the substantial computational costs. This is largely due to the inability to average the initial data volume of 4~TB \citep[after Dysco compression;][]{offringa_compression_2016} before the final imaging, which is constrained by time and bandwidth smearing limits. Current efforts, by revisiting the current data reduction strategies for widefield imaging and improving the software, aim to reduce these computational costs. For deep imaging, we have recently developed techniques to average visibilities at similar baseline coordinates with sidereal visibility averaging when combining multiple observations for deep imaging \citep{2025arXiv250107374D}. This significantly reduces the computational costs required for the final imaging steps, making ultra-deep wide-field imaging in the order of $\mu$Jy~beam$^{-1}$ achievable. Further efforts are being made to run the processing fully automatic, with the aim to facilitate a sub-arcsecond wide-field imaging survey pipeline.

\begin{figure}
    \centering
    \includegraphics[width=0.5\linewidth]{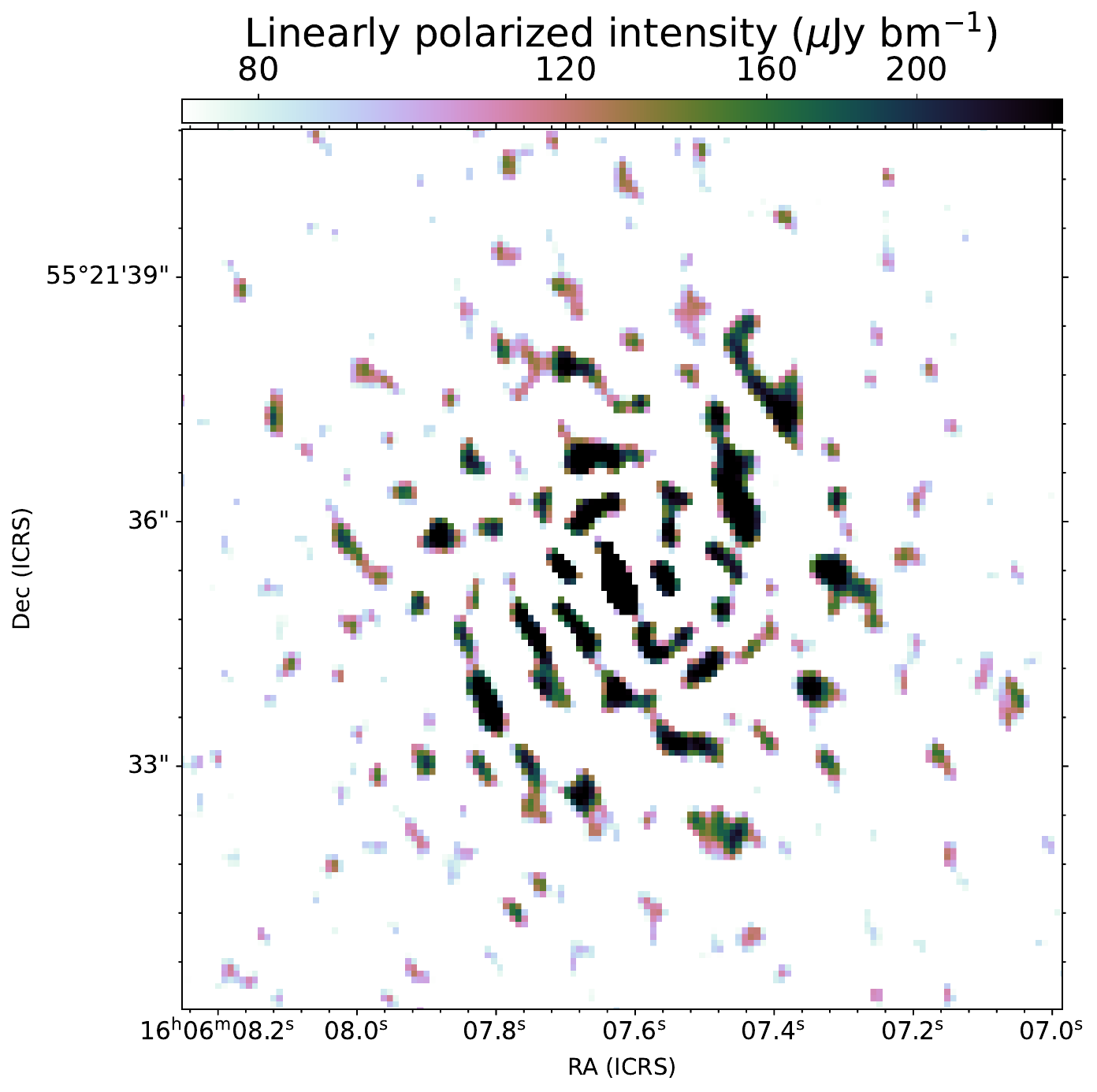}\hspace{-0.3cm}
    \includegraphics[width=0.5\linewidth]{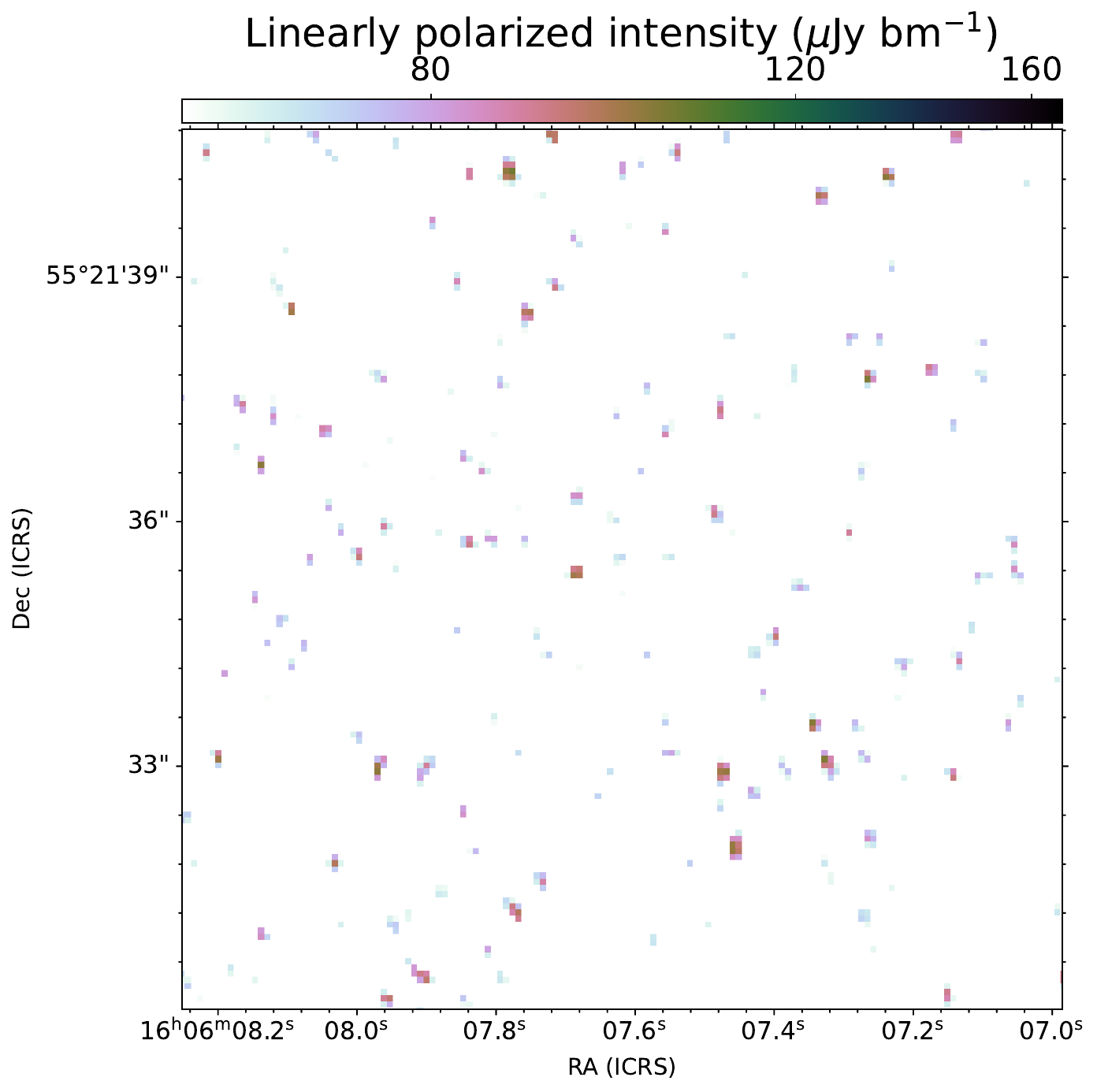}
    \caption{LOFAR Stokes Q maps of an un-polarised delay calibrator in the ELIAS-N1 field, without (left) and with (right) full Jones corrections for the international stations, showing that polarization errors dominating the map on the left are suppressed following our calibration strategy.}
    \label{fig:stokesq_correction}
\end{figure}

\paragraph{Polarisation}
While LOFAR observations (using the Dutch array only) have provided number counts of polarised sources at 144\,MHz \citep[1.24 sources per square degree;][]{piras_lofar_2024}, re-processing the data at 0.\sarc\ 3 allows to reduce the effects of beam depolarization. In addition, the sub-arcsecond resolution enables resolved polarization mapping to distinguish between the jets, lobes, and hotspots of radio galaxies (0.\sarc\ 3 equates to a physical resolution of a few kpc, the typical size of hotspots at that resolution at $z\leqslant0.1$). We are currently testing a strategy for calibrating and imaging of linear polarization in the LOFAR-VLBI pipeline: we adopt a similar method in principle as that of the direction-independent (DI) calibration of the ddf-pipeline (see Section~\ref{subsec:methods}) for the Dutch stations. This includes a correction for ionospheric Faraday Rotation using the RMextract package \citep{2018ascl.soft06024M} which is accurate to about 0.1~rad~m$^{-2}$ \citep[e.g.,][]{2018A&A...613A..58V}.  We produce solutions by self-calibration of the nearby delay calibrator as standard in the LOFAR-VLBI pipeline, but now solving for the full Jones matrix. This translates to finding complex gain corrections for the XX, XY, YX, and YY correlations similar to what is also done for LoTSS processing \citep[for more details see][]{tasse_lofar_2021} for the international stations, see Figure \ref{fig:stokesq_correction}. Here we assume that the delay calibrator is unpolarised. This assumption of an unpolarised delay calibrator can be verified against the LOFAR polarised source catalogue from \cite{2023MNRAS.519.5723O}. We apply the full set of solutions to the target direction, and image the target of interest, creating Stokes Q and U cubes with a large number of channels ($\gtrsim 200$) across the 48\,MHz bandwidth. We then perform RM~Synthesis \citep{brentjens_faraday_2005} to de-rotate the cubes and determine the level of linear polarization at the $RM$ of the rotating medium. We have verified this process on LoTSS data, producing the very first linearly polarised maps of radio sources at 144\,MHz at sub-arcsecond resolution. Using a polarised source in the field, the polarisation angle and $RM$ across multi-epoch LOFAR observations can be aligned allowing deeper studies (van Weeren et al., in prep.), see for example also \cite{2018ASSL..426..159B,2021A&A...648A..12H,2023A&A...674A.119S}. A future pipeline release will include full polarisation calibration, though the level of systematic errors, such as the residual level of instrumental polarisation across the primary beam and residual uncorrected ionospheric Faraday Rotation remains to be studied.\\

\paragraph{The lowest frequencies}
At lower frequencies, work is ongoing to extend all the capabilities that have been successfully developed with HBA observations. Although in principle calibration and processing strategies should be similar, the significant challenges outlined in Section~\ref{subsec:tech} currently limit the extent of ILT LBA observations to a handful of bright 3C sources \citep{groeneveld_pushing_2022}. However, future work could open up new opportunities for ILT LBA observations, as this gives us a second frequency point for spectral studies at high resolution. This is particularly interesting for galaxy clusters, where the emission of radio galaxies would otherwise contaminate the diffuse radio emission permeating galaxy clusters. In addition, high-resolution spectral studies of low-frequency radio emission will allow us to investigate ageing processes in the tails of large radio jets.

The ILT maintains a physical resolution of $<$ 10 kpc across all redshifts even at LBA frequencies, easily allowing for the spectral measurement of multi-component sources on a galaxy-scale. These low frequencies are also the perfect regime to study the effects of synchrotron self-absorption. Turnovers in the spectra, either detected or implied by a lack of detection with the LBA imply highly compact emission, which could be used to more deeply understand the physics for many sources including those still unresolved. 

A major long-term technical goal is the production of a widefield high-resolution LBA image. This is only now becoming feasible with the advance in our techniques for producing these images in \cite{sweijen_high-resolution_2022, de_jong_into_2024} and Escott et al. in prep. The lack of suitable calibrators in multiple directions around the field is the most severe limitation but with the potential to calibrate on faint stacked sources and/or effectively combine multiple datasets with sidereal averaging, we may be close to overcoming these hurdles.
Even further in the future would be advancing LOFAR ILT studies to the shortwave radio band, the lowest frequencies detectable with the LBA antennas (10--30 MHz).
With Dutch baselines, this has recently been shown to be possible \citep{groeneveld_characterization_2024}.
With international stations, however, this would be restrictively challenging at present.
At such low frequencies, differential Faraday rotation caused by  magnetic fields in the ionosphere, and third-order effects become much more significant, especially on the largest baselines.
In addition, LOFAR LBA stations are not designed to operate optimally at such low frequencies, which would significantly limit the ability of the ILT to create sharp images below 30~MHz across a wide field of view.
However, it might be possible to create sharp (resolution below 10 arcsec) images below 30 MHz of bright calibrator sources, during exceptionally favourable ionospheric conditions, which could occur during the solar minimum.

\section{Conclusions}
\label{sec:conclusions}

The past decade has seen massive steps forward, from the first few scientific papers using custom-built calibration strategies, to the development and validation of advanced, modern workflows. From the start, this has been a community-driven effort, with a large percentage of the development coming from PhD students and postdocs. This has resulted in an explosion of science exploitation based on technical developments, including high profile publications in high-impact journals. We are now building towards the next era of LOFAR2.0, and are looking forward to pushing the boundaries with this new facility.

\backmatter

%\bmhead{Supplementary information}

\bmhead{Acknowledgements}

LKM thanks Rosemary Fowler and the Royal Astronomical Society for their support of the Fowler Award for Early Career Achievement. She would also like to thank Adam Deller for his support and guidance in the early years of high resolution and Javier Moldón for his contributions to this area. The authors would like to thank Matthijs van der Wild for his dedicated work producing a Common Workflow Language version of the LOFAR high-resolution imaging pipeline. They thank all LOFAR Long Baseline Working Group members, past and present, for their contributions. 

\section*{Statements and Declarations}

LKM RT, and EE are grateful for support from a UKRI FLF award [MR/Y020405/1]. LKM and RT are grateful for support from STFC grant [ST/V002406/1]. 
This work was supported by the STFC [grant ST/T000244/1].
FS recognizes support from STFC grant [ST/Y004159/1].
JMGHJdJ recognizes support from project CORTEX (NWA.1160.18.316) of research programme NWA-ORC which is (partly) financed by the Dutch Research Council (NWO). 
This work was supported by SURF Cooperative grant no. EINF-6218 and the EGI-ACE project (Horizon 2020 grant no. 101017567). 
VHM acknowledges support of STFC under a grant to the University of Cambridge as part of project reference [ST/Y000447/1].
JP acknowledges support from the CAS–NWO programme for radio astronomy with project number 629.001.024, which is financed by the NWO.
RJvW, CG, and JP acknowledges support from the ERC Starting Grant ClusterWeb 804208.
The authors have no relevant financial or non-financial interests to disclose.
The lead author contributed the conception and design, along with preparation of the introduction, front matter, and back matter content. The other authors contributed with written content and figures. All authors contributed to revisions for subsequent drafts. All authors read and approved the final manuscript.

\bibliography{references}

\end{document}